\documentclass[traditabstract]{aa}
\pdfoutput=1 \pdfoutput=1 
             
\usepackage{epsfig,caption,rotating,color}
\usepackage{multirow}
\usepackage{graphics}
\usepackage[latin1]{inputenc}
\usepackage{natbib}
\bibpunct{(}{)}{;}{a}{}{,}

\begin{document}

\title{\bf 
The  Extended Baryon Oscillation Spectroscopic Survey: Variability Selection and Quasar Luminosity Function 
}
\author{N. Palanque-Delabrouille\inst{1}, Ch. Magneville \inst{1}, Ch. Y\`eche\inst{1}, I. P\^aris\inst{2}, P. Petitjean\inst{3}, E. Burtin\inst{1}, K. Dawson\inst{4}, 
I.~McGreer\inst{5}, A. D. Myers\inst{6}, G. Rossi\inst{7}, D. Schlegel\inst{8}, D. Schneider\inst{9,10}, A. Streblyanska\inst{11,12}, J. Tinker\inst{13}  } 

\institute{CEA, Centre de Saclay, Irfu/SPP,  F-91191 Gif-sur-Yvette, France
\and INAF - Osservatorio Astronomico di Trieste, Via G. B. Tiepolo 11, I-34131 Trieste, IT 
\and UPMC-CNRS, UMR7095, Institut d?Astrophysique de Paris, F-75014, Paris, France 
\and Department of Physics and Astronomy, University of Utah, Salt Lake City, UT 84112, USA 
\and Steward Observatory, University of Arizona, 933 North Cherry Avenue, Tucson, AZ 85721, USA 
\and Department of Physics and Astronomy, University of Wyoming, Laramie, WY 82071, USA  
\and Department of Astronomy and Space Science, Sejong University, Seoul, 143-747, Korea 
\and Lawrence Berkeley National Laboratory, 1 Cyclotron Road, Berkeley, CA94720, USA 
\and Department of Astronomy and Astrophysics, 525 Davey Laboratory, The Pennsylvania State University, University Park, PA 16802, USA 
\and Institute for Gravitation and the Cosmos, The Pennsylvania State University, University Park, PA 16802, USA
\and Instituto de Astrofisica de Canarias (IAC), E-38200 La Laguna, Tenerife, Spain 
\and  Universidad de La Laguna (ULL), Dept. Astrofisica, E-38206 La Laguna, Tenerife, Spain 
\and Center for Cosmology and Particle Physics, New York University, New York, NY 10003, US  
}
\date{Received xx; accepted xx}
\authorrunning{N. Palanque-Delabrouille et al.}
\titlerunning{Quasar Luminosity Function  }
\abstract{
The extended Baryon Oscillation Spectroscopic Survey of the  Sloan Digital Sky Survey (SDSS-IV/eBOSS) has an extensive quasar program that combines several selection methods. 
Among these, the photometric variability technique provides highly uniform samples, unaffected by the  redshift bias of traditional  optical-color selections,  when  $z= 2.7 - 3.5$ quasars cross  the stellar locus or when host galaxy light affects quasar colors at $z < 0.9$.  
Here, we present the variability selection of quasars in eBOSS, focusing on a specific program  that led to a sample of 13,876 quasars to $g_{\rm dered}=22.5$ over a 94.5~deg$^2$ region in Stripe 82, an areal density 1.5 times higher than over the rest of the eBOSS footprint.

We use these variability-selected data to provide  a new measurement of the quasar luminosity function (QLF) in the redshift range $0.68<z<4.0$. Our sample is more dense, and reaches deeper than those used in previous studies of the QLF, and is among the largest ones.
At the faint end, our QLF extends to $M_g(z\!=\!2)=-21.80$ at low redshift and to $M_g(z\!=\!2)=-26.20$ at $z\sim 4$. 
We fit the QLF using two independent double-power-law models with ten free parameters each. The first model  is  a pure luminosity-function evolution (PLE) with bright-end and faint-end slopes allowed to be different on either side of $z=2.2$. The other is a simple PLE at $z<2.2$, combined with a model that comprises both luminosity and density evolution (LEDE) at $z>2.2$. Both models are constrained to be continuous at $z=2.2$. They  present a flattening of the bright-end slope at large redshift. The LEDE model indicates a reduction of the break density with increasing redshift, but the evolution of the break magnitude depends on the parameterization. The models  are in excellent accord, predicting quasar counts that agree within 0.3\% (resp., 1.1\%) to $g<22.5$ (resp., $g<23$). The models are also in good agreement  over the entire redshift range with models from previous studies.

 }
\maketitle

\section{Introduction}
\label{sec:intro}
Quasars have become a key ingredient in our understanding of cosmology and galaxy evolution.  Being among the most luminous extragalactic sources, they  have become a mainstay of cosmological  surveys such as the 2dF Quasar Redshift Survey~\cite[2QZ;][]{Croom2001} and the Sloan Digital sky Survey~\citep[SDSS;][]{York2000}, where they are the source of choice to study  large-scale structures at high redshift. Quasars  can be used as direct tracers of dark matter in the redshift range $0.9<z<2.1$ where they are present at sufficiently high density, and as background beacons to illuminate the intergalactic medium at higher redshift, where the cosmological information is produced by the foreground neutral-hydrogen absorption systems that form the Lyman-$\alpha$ forest. As part of the third-generation of the Sloan Digital Sky Survey  \citep[SDSS-III;][]{Eisenstein2011}, the Baryon Acoustic Oscillation Survey \citep[BOSS;][]{Dawson2013} measured the spectrum of about 300.000 quasars,   $180.000$ of which are at $z>2.15$, to a limiting magnitude of $g\sim 22$. As part of SDSS-IV, the extended Baryon Oscillation Spectroscopic Survey \citep[eBOSS;][]{Dawson2015, Tinker2015} is aiming to more than quadruple the number of known quasars over redshifts of $0.9 < z < 2.2$ to $g\sim22$, in addition to targeting new quasars at $z > 2.2$. The next-generation Dark Energy Spectroscopic  Instrument \citep[DESI, previously named BigBOSS;][]{Schlegel2011} is designed to obtain spectra of 
more than two million quasars, reaching limiting magnitudes $g\sim 23$. This new challenge requires, as a first step, a good knowledge of the quasar luminosity function (QLF) in order to determine the expected number count for quasars, and optimize the distribution of fibers among the various cosmological probes. 

In the past two decades, with the advent of large quasar surveys, the number of known quasars has increased by over a factor 20,  triggering significant effort to measure  the QLF \citep[see][for an overview of  recent determinations]{Ross2013}.  Nevertheless, the measurement of the QLF over $2<z<4$,  where the number density of quasars starts to decline and  their selection with traditional color-based algorithms is less efficient, remains challenging, especially at the faint end. This situation arises  because the broad-band colors of $z\sim 2.7$ and $z\sim 3.5$ quasars are, respectively, very similar to those of A-F and K stars \citep{Fan1999, Fan2001, Richards2002, Ross2012}.  
The density of $z < 0.9$ quasars is not well-characterized either as 
host galaxy light can significantly affect the colors of faint quasars.
To circumvent these difficulties, \cite{Palanque2011} developed a selection algorithm relying on the time variability of quasar fluxes.  This technique was demonstrated to increase by 20 to 30\% the density of identified quasars, and to effectively recover additional quasars in the redshift range $2.5<z<3.5$. It was applied to measure the QLF over the redshift range $0.68<z<4.0$ \citep{Palanque2013a}, using a sample of quasars that was found to be 80\% complete to $g=20$, and still 50\% complete at $g=22.5$. Despite its limited statistics of 1877 quasars, this study yielded competitive results which have been used to estimate quasar counts for several ongoing large-area surveys.  

The QLF can only be improved with a well-controlled quasar sample  of much larger size. In the near-term, eBOSS is the most ambitious survey verifying this requirement. Designed to measure the scale of the Baryon Acoustic Oscillations, or BAO~\citep{Eisenstein2005}, at the 2\% level in the yet unexplored  $0.9<z<2.2$ regime, eBOSS plans to target and spectroscopically identify at least 500,000 quasars in this redshift range, including quasars already confirmed with SDSS-I/II. At $z>2.1$, eBOSS will complement previous studies from BOSS~\citep{Slosar2013,Busca2012,Delubac2015} and provide a  measurement of the BAO feature at 1.5\%, using a sample of 75,000 quasars that had not previously been identified. In addition, eBOSS  has conducted an extensive search for quasars at all redshifts in a $120~{\rm deg}^2$  area where unique time-domain photometry from SDSS is available. Because this region allows a highly complete selection of quasars with minimal completeness corrections, it is ideal  for QLF studies, and is the focus of the present work. We improve upon previous QLF studies, such as \citet{Croom2009}, \citet{Palanque2013a} and \citet{Ross2013}, in  terms of the size of the quasar sample used for the measurement, in  depth, and in  redshift homogeneity of the target selection. For the quasar luminosity function, we build on earlier semi-empirical models  \citep[e.g., ][]{Schmidt1983, Koo1988,
Boyle1988, Boyle2000, Croom2004, Richards2005, Hasinger2005, Richards2006, Hopkins2007} 
such as Pure Luminosity Evolution (PLE), models that evolve exponentially with look-back time,
Luminosity Dependent Density Evolution (LDDE) and Luminosity Evolution + Density Evolution (LEDE).

In this paper, we present a sample of 13,876 quasars, selected by eBOSS over a 94.5 ${\rm deg}^2$ region with a technique  relying upon quasar time-domain variability. For this study, we take advantage of spectroscopy conducted by eBOSS of the part of the SDSS Southern equatorial stripe, hereafter referred to as Stripe 82~\citep{Stoughton2002},  where 50 to 100 epochs of imaging are available over a time period of about 10 years. The variability technique used here is a robust, efficient and well-understood method  whose completeness can be readily evaluated using an independent control sample. With this strategy, all completeness corrections can be derived from the data, without requiring any model of quasar light curves or colors. 

The outline of the paper is as follows. In Sec.~\ref{sec:eBOSS_var} we present the variability programs in eBOSS.  In Sec.~\ref{sec:data}, we describe the imaging data, provide the details of the selection of the targets for this study and present the resulting spectroscopic data. In Sec.~\ref{sec:counts}, we give the raw quasar number counts,  explain the computation of the completeness corrections, and  derive the completeness-corrected number counts. Finally, in Sec.~\ref{sec:LF}, we derive the QLF from our data. The present analysis refers extensively to our previous works on quasar variability. To simplify the presentation and make it easier for the reader to identify any references to these earlier papers, we will henceforth refer to our paper demonstrating the use of time-domain variability for quasar selection as Paper~Var~\citep{Palanque2011}, and to our paper presenting our previous measurement of the QLF as Paper~LF~\citep{Palanque2013a}.

\section{Time-domain quasar selection with eBOSS}
\label{sec:eBOSS_var}

Data for SDSS-IV/eBOSS is taken with the 2.5-meter Sloan Foundation Telescope~\citep{Gunn2006}, using the same spectrograph and data reduction pipeline as for SDSS-III/BOSS~\citep{Bolton2012, Smee2013, Dawson2013}. The eBOSS survey~\citep{Dawson2015} includes  an extensive quasar program~\citep{Myers2015}. A CORE selection will provide a homogeneous sample of at least 69 deg$^{-2}~0.9<z<2.2$ quasars, and  the combination of several techniques will increase the sample of $z>2.1$ quasars by more than $\sim 7~{\rm deg}^{-2}$ compared to BOSS. The majority of the  quasars at $z>2.1$ are obtained either from the  CORE selection, which is not strictly limited to $z<2.2$ and provides of order $6~{\rm deg}^{-2}$, or from a selection based on quasar variability that provides another $\sim 3~{\rm deg}^{-2}$ quasars. 

The use of time-domain photometric measurements to exploit quasars' intrinsic variability has been demonstrated during the course of the BOSS survey in Paper~Var and Paper~LF. In the context of eBOSS, variability selection of quasars is performed over 90\% of the survey footprint  using time-domain data from the Palomar Transient Factory~\citep[PTF: ][]{Rau09,Law09}.  Details on the variability selection from PTF data are available in \citet{Myers2015}. 
Over Stripe 82, however, the SDSS provides data that are both deeper and with a longer lever-arm in time than  PTF. In this region, we therefore replace the PTF selection by  a dedicated program (VAR\_S82) based on a variability selection of quasars from SDSS photometry. 

In Tab. ~\ref{tab:eBOSS_TSbits}, we list the  three  selection methods dedicated to quasars, with their eBOSS  targeting bit names,  numerical equivalents, and  average target density  over Stripe 82 for  CORE and  VAR\_S82, and  over the eBOSS footprint where PTF data are used, thus outside Stripe 82, for  PTF.   We also provide the density of quasars already identified spectroscopically   in Stripe 82 (hereafter referred to as the `known' quasars), which is higher than over the rest of the eBOSS footprint because of several BOSS programs dedicated to quasar selection in Stripe 82. The listed density for CORE is  after removal of the overlap with  known quasars, and  the density for both  PTF and VAR\_S82 are given after removal of the overlap with both known and  CORE samples.
 
 \begin{table}[htb]
\begin{center}
\begin{tabular}{lcr}
\hline\hline
Bit & Name & Density (deg$^{-2}$)\\
\hline
- & {\it Known quasars} & 80 \\
$2^{10}$& QSO\_EBOSS\_CORE& 60 \\
$2^{11}$ & QSO\_PTF & 20  \\
$2^{9}$ & QSO\_VAR\_S82 & 50  \\
  \hline
  \end{tabular}
  \end{center}
\caption{eBOSS quasar targeting bits and average targeting densities in Stripe 82, except for PTF. The CORE density is that after removal of the overlap with  known quasars. The PTF and VAR\_S82 densities are given after removal of the overlap with both CORE and known quasars.}
\label{tab:eBOSS_TSbits}
\end{table}

All BOSS quasar targets were visually inspected to be classified as star, galaxy or quasar~\citep{Paris2012, Paris2014}. This procedure  is no longer possible in eBOSS where the density of quasar targets is increased by at least a factor 4 (about a factor 5 in specific regions such as Stripe 82). Studies on the pipeline performance allowed an improvement of the consistency  between pipeline and visual-inspection classifications, and thus significantly reduced the need for visual inspection~\citep[cf. ][for details on how spectra are selected for visual inspection]{Dawson2015}. We  identified the types of failures that could not be systematically associated with any given class. The remaining failures are flagged as requiring visual inspection. As a result of these improvements, the fraction of visually-inspected quasar targets, including both those that were identified as needing an inspection and those belonging to the few random plates that were visually inspected for evaluation of the pipeline performance, is of order 8\% on average over the eBOSS footprint. For the  QSO\_VAR\_S82 targets, however, this fraction rises to $\sim$17\% due to the fainter brightness on average of the selected objects.

\begin{figure}[htbp]
\begin{center}
\epsfig{figure= 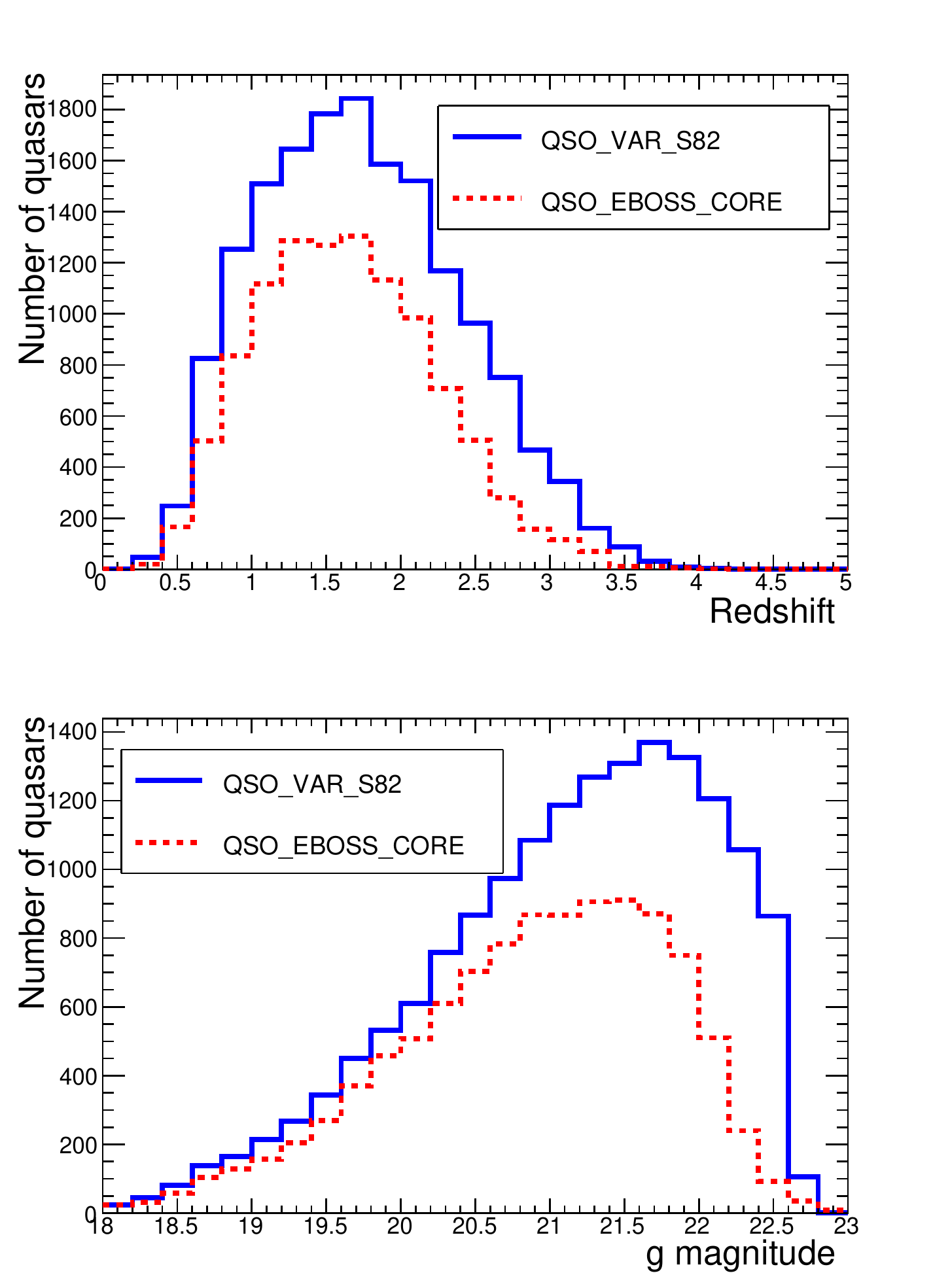,width = \linewidth} 
\caption{\it Redshift and magnitude distributions of the quasars selected over Stripe 82 with the QSO\_EBOSS\_CORE (10481 objects) or the QSO\_VAR\_S82 (16243 objects) flag. } 
\label{fig:var_vs_core}
\end{center}
\end{figure}

For all the objects, the eBOSS pipeline encodes the spectrum classification into the CLASS\_AUTO flag, and the redshift, when relevant, into Z\_AUTO. When the spectrum was  visually inspected, two additional flags are set: CLASS\_PERSON, which encodes the classification, and Z\_CONF\_PERSON, which encodes  the confidence on the redshift estimate~\citep{Paris2014}.  We define a spectrum ``uberclass'' as follows.
If a visual inspection was done and  led to a clear identification (Z\_CONF\_PERSON$\ge 2$), then the object uberclass is set equal to CLASS\_PERSON (i.e., Star, Quasar or Galaxy).  If  visual inspection did not lead to a clear identification, 
the  uberclass is set to `Inconclusive'. In the absence of visual inspection, the uberclass is set equal to CLASS\_AUTO, which can also be Inconclusive. 
We define our  sample of quasars as the set of targets with uberclass equal to Quasar. The quasar redshift is set equal  to the visual inspection redshift if the latter is  available, and to Z\_AUTO otherwise.

We illustrate in Fig.~\ref{fig:var_vs_core} the magnitude and redshift distributions of the quasars selected in Stripe 82 with the QSO\_EBOSS\_CORE or the QSO\_VAR\_S82 flag, including  the known quasars (and the overlap with CORE in the variability sample). A large fraction of the quasars are common to both selections. The QSO\_VAR\_S82 sample, however,  contains about 1.6 times more quasars than the QSO\_EBOSS\_CORE sample. The origin of this improvement is 2-fold. Part of it is due to the fact that the QSO\_VAR\_S82 sample is selected from about 50 epochs of photometry, instead of a single epoch for the QSO\_EBOSS\_CORE sample (which is done to ensure the uniformity with the rest of the eBOSS footprint). The other part  of the improvement comes from the different selection techniques: at identical depth for the input photometry, using for instance 50-epoch coadded images to measure object colors and 50 individual epochs of imaging to measure  variability criteria, we have shown in Paper Var that the variability selection selects 30\% more quasars than the CORE selection for the same total number of targets.

The QSO\_VAR\_S82 sample significantly increases the  completeness at all redshifts and magnitudes, and  in particular  at the faint end ($g\sim$ 22--22.5). 
In the rest of this paper, we focus on the targets  with the QSO\_VAR\_S82 bit set.


\section{Data and target selection}
\label{sec:data}

This section provides an overview of the control sample of quasars used throughout this study. We  present the imaging data from which the targets are selected, describe the selection algorithm and discuss the spectroscopic survey. The magnitudes of the sources are denoted $u$, $g$, $r$, $i$ and $z$ when referring  to observed magnitudes, and $u_{\rm dered}$, $g_{\rm dered}$, $r_{\rm dered}$, $i_{\rm dered}$ and $z_{\rm dered}$ when referring to magnitudes corrected for Galactic extinction using the extinctions from the maps of~\citet{Schlegel1998}. The bands correspond to the SDSS filters~\citep{Fukugita1996, Doi2010}.

\subsection{Control sample}\label{subsec:controlsample}

The completeness corrections related to the criteria used to select the targets are determined using a control sample of 4555 spectroscopically-confirmed quasars that were selected in Stripe 82 independently of any variability criterion. This prevents our control sample from being biased towards sources that exhibit a high quasar-like variability. Such quasars would indeed have led to over-optimistic completeness estimates for our selection algorithm. 

The control sample is built from  the 2dF quasar catalog 
\citep{Croom2004}, the 2dF-SDSS LRG and Quasar Survey 2SLAQ ~\citep{Croom2009}, the SDSS-DR7 quasar catalog~\citep{Schneider2010} and  BOSS observations through August 2010~\citep{Ahn2012,Paris2012}. 
These  catalogs were  obtained from pure color selections (cf. for instance \cite{Richards2002} for DR7 and \cite{Ross2012} for BOSS).
BOSS observations taken after Summer 2010 on Stripe82 had contributions from a variability selection and were therefore discarded from the control sample. 

Color information comes from flux ratios, thus not sensitive to absolute fluxes, while variability information comes from variations in absolute fluxes. Furthermore, time variations are seen to be synchronous in different bands, thus not affecting the source colors. Color and variability selections are therefore complementary, as was already shown in Paper~Var, with no obvious correlations between the photometric and time-domain characteristics of quasars. The only source of correlations could come from the image depth, but we take this into account by computing all corrections as a function of source magnitude. 
We thus expect no measurable bias in our completeness estimates from the use of this control sample.

\subsection{Imaging data and target selection}\label{subsec:ts}

The selection of the targets for this study relies heavily on the variability selection described in Paper~Var where all the details can be found; we therefore only summarize the main steps here. 

\begin{figure}[htbp]
\begin{center}
\epsfig{figure= 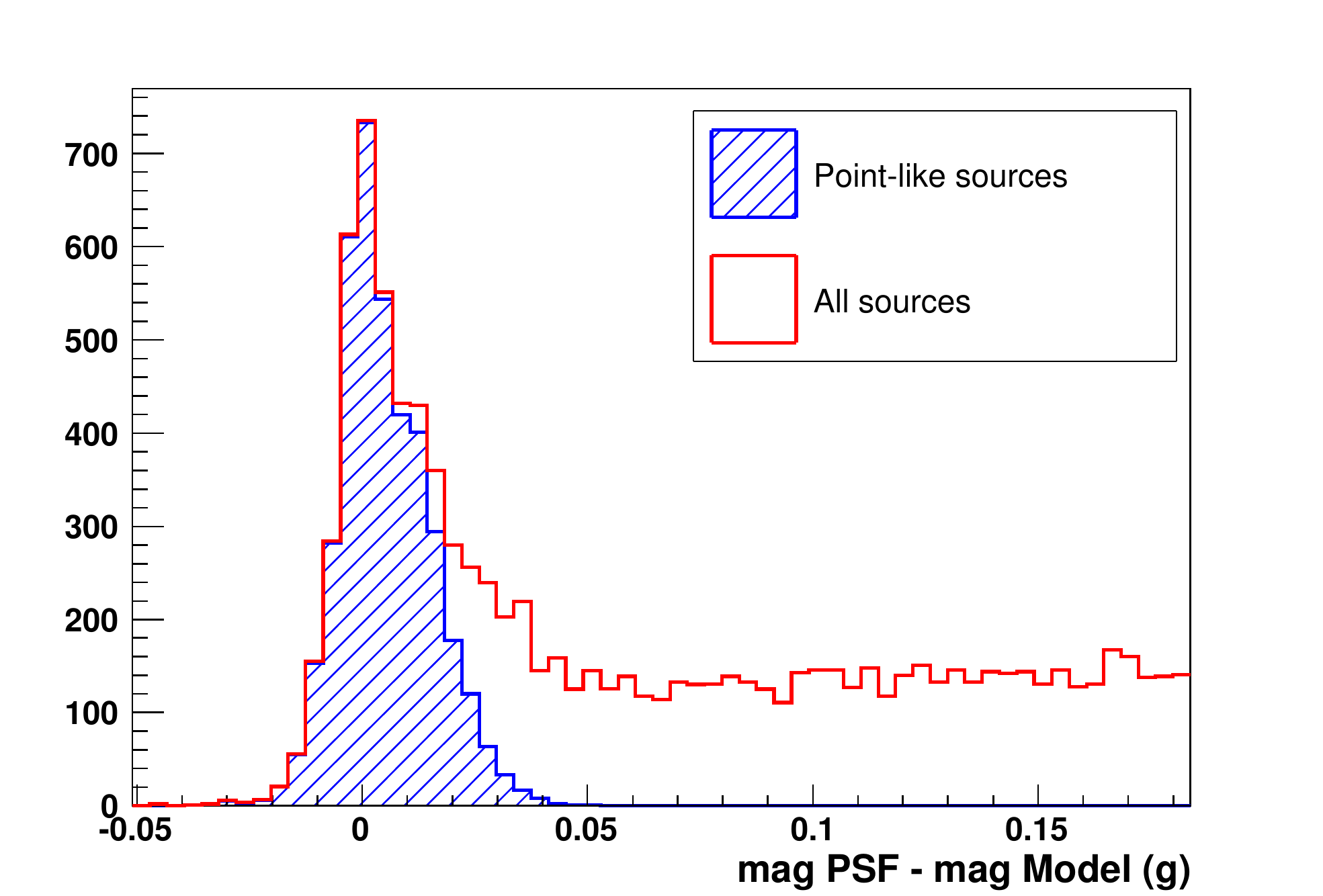,width = \linewidth} 
\caption{\it Difference between PSF and model magnitudes, used as morphology indicator, for random objects in Stripe 82. Point-like sources in the co-added images, overlaid in blue, have a small magnitude difference. Quasar selection requires  magnitude differences of at most 0.05, relaxed to 0.1 for objects with high significance of quasar-like variability (to  recover of low-$z$ quasars where the host galaxy can  make the source appear  extended). } 
\label{fig:morphology}
\end{center}
\begin{center}
\epsfig{figure= 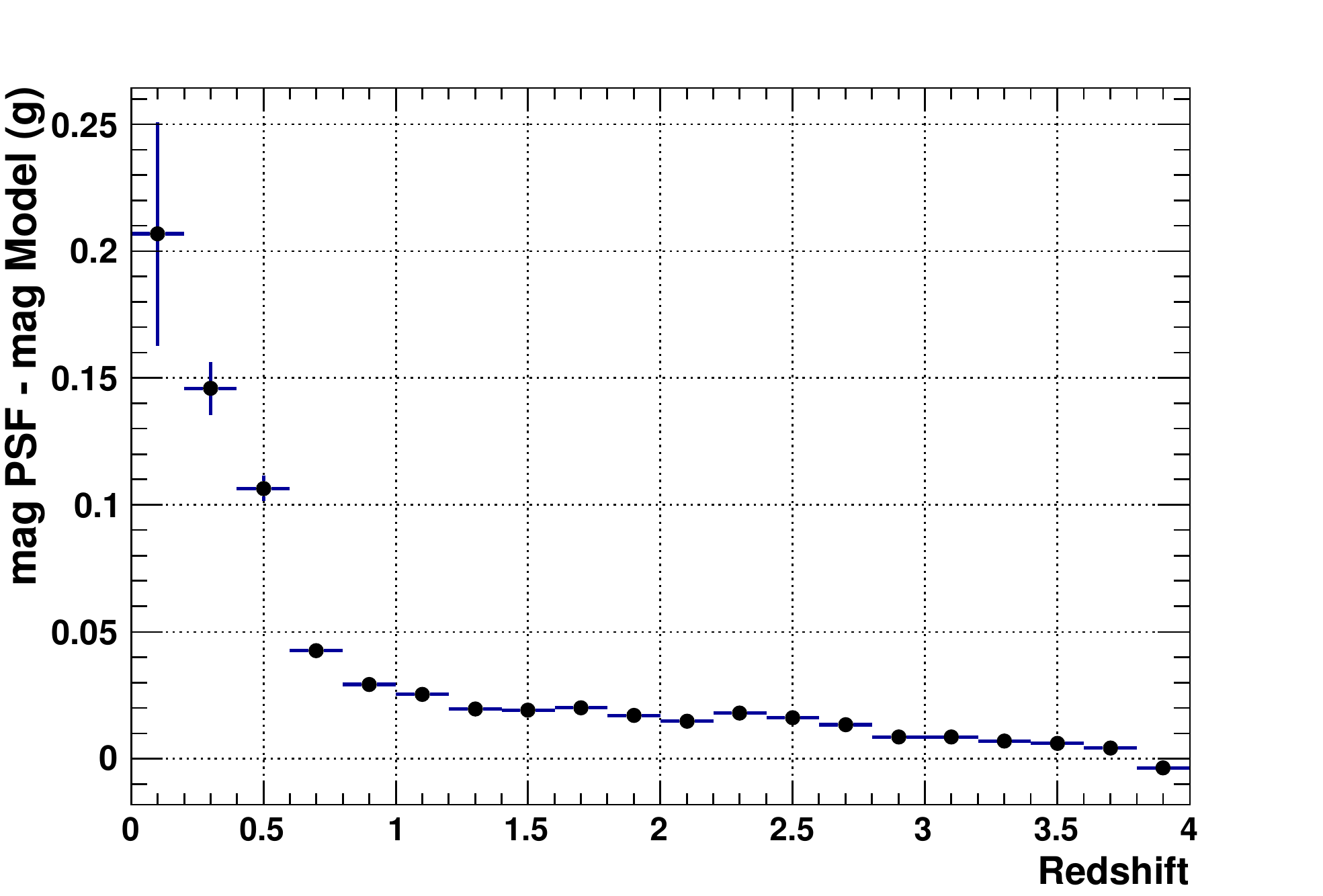,width = \linewidth} 
\caption{\it Morphology indicator as a function of redshift for quasars. The cut at 0.05 includes most quasars at $z>0.6$, but rejects most lower-redhift quasars whose host galaxy is detectable in co-added images. } 
\label{fig:qso_morphology}
\end{center}
\end{figure}

The initial source list is determined  from the co-addition of single-epoch SDSS images~\citep{Annis2014} in Stripe 82, from which we take the source magnitude (in SDSS $u$, $g$, $r$, $i$ and $z$ bands) and morphology. As morphology indicator, we use a continuous variable defined as 
\begin{equation}
m_{\rm diff} = m_{\rm PSF} (g) - m_{\rm model} (g)~,
\end{equation}
where $m_{\rm PSF} (g)$ and $m_{\rm model} (g)$ are the magnitudes of the source, measured in the $g$ band, obtained from a PSF fit (valid for unresolved objects) or a model fit (more appropriate for extended objects, where model can be a de Vaucouleurs or an exponential shape, for instance), respectively. 
As shown in Fig.~\ref{fig:morphology}, $m_{\rm diff}$ peaks near 0 with a standard deviation of 0.01 for point-like sources, and  extends from 0.01 to values beyond 0.3 for extended sources. Because the emitting region of a quasar, too small to be resolved, outshines the host galaxy by a large factor, a quasar generally appears as a stellar-like source. Only for  the nearest quasars (redshift $z<0.9$ at most) can the host galaxy be detectable in co-added images, making the source appear extended in those cases (see Sec.~2.4 of Paper~LF or a quantitative study of the effect). The average morphology indicator as a function of redshift for the quasars of the control sample  is displayed in Fig.~\ref{fig:qso_morphology}. It is at the level of 0.2 or more as the redshift approaches 0, has a value of about 0.1 at  $z\sim 0.5$, and is below 0.05 for redshifts $z>0.6$.
At  $z> 2.5$, the small decrease in $m_{\rm diff}$ with increasing redshift is correlated with the increasing  average magnitude of the quasars. The larger the photometric errors, the less the  PSF and model fits tend to  capture a difference in spatial extension of the source, and hence the closer on average is $m_{\rm diff}$  to zero.  
 We apply a cut on morphology (upper bound on $m_{\rm diff}$, cf. details below) to reject galaxies. 

To reduce the stellar contamination, we apply a  loose color cut by requiring that $c_3<1-  0.33\times c_1$, where  $c_1$ and $c_3$ are defined as in~\cite{Fan1999} by 
\begin{eqnarray}
c_1 &=& 0.95(u-g)+0.31(g-r)+0.11(r-i) \, ,\nonumber \\
c_3 &=& -0.39(u-g)+0.79(g-r)+0.47(r-i)\, .
\label{eq:c1c3} \end{eqnarray} 
This is the same criterion as was used  in Paper~LF, where we had estimated that close to 100\% of $z<2.2$ and 98\% of $z>2.2$ known quasars (i.e., quasars from the control sample of Sec.~\ref{subsec:controlsample}) passed that condition. The completeness of this color cut is included in the selection efficiency $\epsilon_{\rm sel}$ described in Sec.~\ref{sec:completenesscor}. 

The main selection is based upon a criterion measuring the  variability with time of the source. The lightcurves of our sources contain, on average, 52 SDSS individual epochs spread over 7 years. They are used to compute two sets of parameters that characterize the source variability:\\
 - the  $\chi^2$ of the fit of the lightcurve in each of the $ugriz$ filters by a constant $\overline{m}$:
$\chi^2 = \sum_i \left[  (m_i - \overline{m}) / \sigma_i \right] ^2$, where the sum runs over all observations $i$, \\
- two parameters, an amplitude $A$ and a power $\gamma$ as introduced by \cite{Schmidt2010}, that characterize the variability structure function $\mathcal V(\Delta t_{\rm ij})$, i.e.,  the change in magnitude $\Delta m_{\rm ij}$ as a function of time lag $\Delta t_{\rm ij}$ for any pair $ij$ of observations: 
$ \mathcal V(\Delta t_{\rm ij}) = |\Delta m_{\rm i,j}| - (\sigma_{\rm i}^2 + \sigma_{\rm j}^2)^{1/2} 
=A\times (\Delta t_{\rm ij})^\gamma$. Because quasars have similar time variations in different bands, we reduce the uncertainty on variability parameters by fitting simultaneously the $g$, $r$ and $i$ bands (those least affected by noise and observational limitations) for a common $\gamma$ and independent amplitudes $A_g$, $A_r$ and $A_i$. 

Variable objects, whether quasars or  stars, are expected to have large $\chi^2$'s, thus allowing a distinction between variable and non-variable targets. The structure function parameters $A$ and $\gamma$ can discriminate between these two classes of variable objects: quasars tend to have both large $A$ and large $\gamma$, due to magnitude changes that increase with time, while variable stars (such as pulsating or eclipsing binaries) can have large $A$ but usually $\gamma$ near 0. 

For each source, a neural network  combines  the five $\chi^2$, the power $\gamma$ and the amplitudes $A_g$, $A_r$ and $A_i$, to produce an estimate of quasar-like variability.   The training of the NN was done using a large sample of 13~063 spectroscopically confirmed quasars with redshifts in $0.05\le z\le 5.0$ and magnitudes in the range $18\le g\le 23$, and a star sample consisting in 2~609 objects spectroscopically confirmed as stars in the course of the BOSS project. The two samples are located in Stripe 82, and thus have identical time sampling characteristics as the present data.  An output $v_{\rm NN}$ of the neural network near $0$ designates non-varying objects, as is the case for the vast majority of stars, while an output near $1$ indicates lightcurves exhibiting quasar-like variability. \\

A source is selected according to its quasar-like variability ($v_{\rm NN}$) and  morphology ($m_{\rm diff}$). A loose morphology cut ($m_{\rm diff}<0.1$) is applied if the variability indicator is high ($v_{\rm NN}>0.85$), a strict morphology cut ($m_{\rm diff}<0.05$) is applied in case of a lower variability indicator  ($0.50<v_{\rm NN}<0.81$), and for  values of  $v_{\rm NN}$ intermediate between 0.81 and 0.85, the threshold on $m_{\rm diff}$ is gradually evolved between the two extreme values of 0.05 and 0.1. Even the tightest bound (0.05)  fully encompasses the range of potential values for point-like sources.

We restrict the study to sources with $g<22.8$ and $g_{\rm dered}<22.5$. The average $g$ Galactic extinction over the observed zone of Stripe 82 is 0.12, so both limits are comparable. With this magnitude limit, the selection described above leads to  a sample density of  175~${\rm deg}^{-2}$ targets. 
Removing targets that already have a spectroscopic identification from previous observations reduces the sample to about  95~${\rm deg}^{-2}$ targets.  Further removing the overlap with the CORE sample (bit QSO\_EBOSS\_CORE) yields the target density of $50~{\rm deg}^{-2}$ indicated in Tab.~\ref{tab:eBOSS_TSbits}.

\subsection{Spectroscopic data}\label{subsec:data}

\begin{figure}[htbp]
\begin{center}
\epsfig{figure= 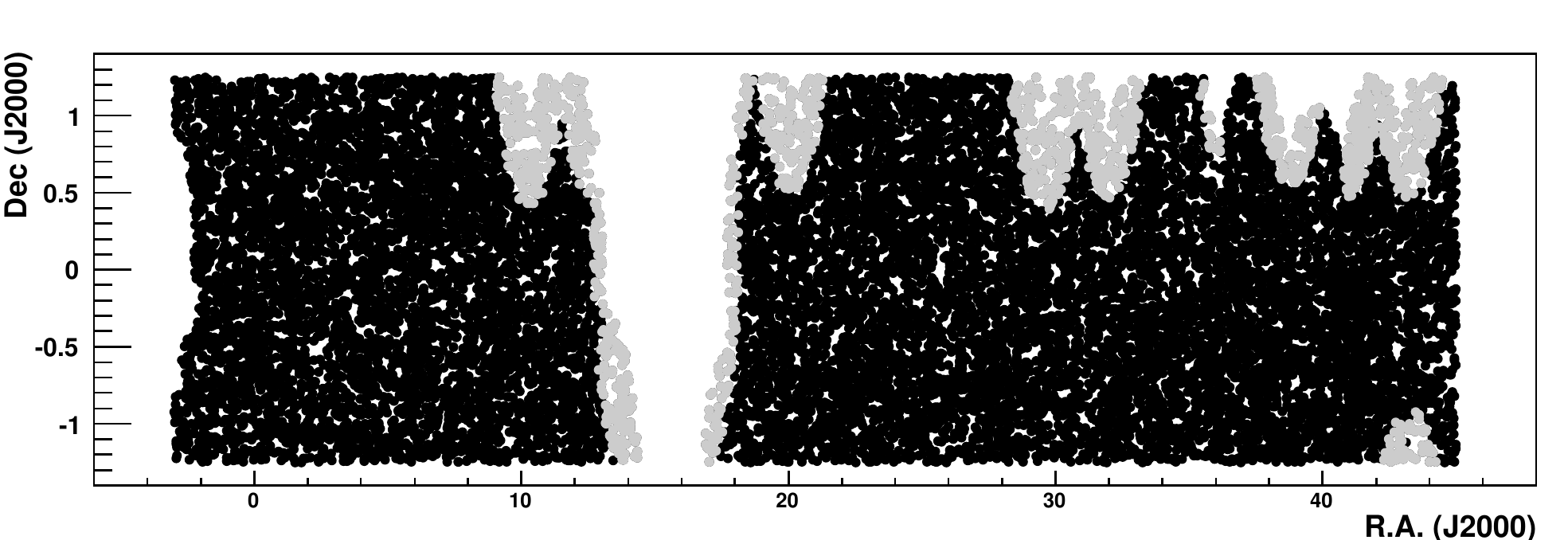,width = \linewidth} 
\caption{\it Footprint of the first year of eBOSS observations used for this study (aspect ratio is not 1:1).  Incomplete regions are shown in gray,  fully observed ones in black. } 
\label{fig:footprint}
\end{center}
\end{figure}

The eBOSS footprint overlaps Stripe 82 over  a total of 120~${\rm deg}^2$ delimited by $-3^{\circ}<\alpha_{\rm J2000} <45^{\circ}$ and $-1.25^{\circ}<\delta_{\rm J2000}<1.25^\circ$. The first year of eBOSS observations led to the coverage illustrated in Fig.~\ref{fig:footprint}: dots indicate the position of quasars in plates that have been observed by eBOSS. Some outskirt regions (shown in gray in the figure) have less than 100\% completeness because of overlapping plates that are yet to be observed at the time of this work. In the present analysis, we  restrict ourselves to the region that has 100\% completeness (in black in the figure). Its total area is $94.5~{\rm deg}^2$.

In the course of the BOSS survey, several ancillary projects have covered  parts or all of Stripe 82, either to test target selection techniques (Paper~Var from quasars targeted with BOSS chunk 11 over all Stripe 82),  or as pilot programs for eBOSS and DESI~\citep{Ross2012, Dawson2013,  2015arXiv150100963A}. In particular,  
three programs aimed to  provide an exhaustive census of  quasars to  at least $g_{\rm dered}\sim 22.5$ in  Stripe 82. The first one, conducted jointly in BOSS (chunk 21,  target bit QSO\_VAR\_FPG) and  on the Multiple Mirror Telescope, was done in the 
region delimited by $317^\circ<\alpha_{\rm J2000}<330^\circ$. It provided a total of nearly 1800 quasars at a mean magnitude $\langle g_{\rm dered} \rangle= 21.1$  and led  to the QLF paper mentioned previously (Paper~LF). The other two programs were conducted in the region of Stripe 82
delimited by $36^\circ<\alpha_{\rm J2000}<42^\circ$ where Galactic extinction is low \cite[c.f.,  DR12 release of SDSS-III:][] {2015arXiv150100963A}: one program (BOSS chunk 205, target bit QSO\_VAR\_LF) identified about 1600 new quasars  to $g_{\rm dered}\sim22.5$, while the second (BOSS chunk 218, target bit QSO\_DEEP),  aimed at identifying fainter targets  and provided an additional 363 quasars at $\langle g_{\rm dered} \rangle= 22.6$. The QSO\_VAR\_LF and the QSO\_DEEP programs more than doubled  the density  of known quasars given in Tab.~\ref{tab:eBOSS_TSbits} for  the rest of Stripe 82. 
The $36^\circ<\alpha_{\rm J2000}<42^\circ$ region of Stripe 82 where a deep quasar sample is available is used in this work to cross-check  the completeness-corrected counts computed in Sec.~\ref{sec:counts}.


\section{Quasar number counts}
\label{sec:counts}

We here compute the completeness corrections that affect our sample, whether they are related to the analysis technique or to  observational constraints. We   present  raw number counts derived from the observation of our targets, and compute  corrected number counts  used in Sec.~\ref{sec:LF} to derive a quasar luminosity function. As mentioned above, we only give counts within the fully observed zone (black area in Fig.~\ref{fig:footprint}), of area 94.5 deg$^2$.

\subsection{Completeness corrections}\label{sec:completenesscor}

We derive the spectroscopy-related completeness corrections  from the data themselves, and the selection-related corrections from the application of the same selection cuts to the control sample of quasars. We describe below the different contributions. 

\subsubsection*{Morphology completeness,  $\epsilon_{\rm morph}(z), g$: }

As explained in Sec.~\ref{subsec:ts}, we only select targets among the  sources that pass the morphology cut $m_{\rm diff} < 0.1$. Therefore, sources that are more extended than allowed by this cut are not considered as possible candidates.  Nevertheless,  some  low-redshift quasars for which the host galaxy is resolved can fail this criterion (c.f., Fig.~\ref{fig:morphology}). We compute the correction related to this incompleteness by considering the fraction   of  the quasars in our control sample that have $m_{\rm diff} > 0.1$, as a function of redshift and magnitude.  
This procedure  slightly underestimates the effect, since the control sample is dominantly consisting of objects that were already selected to be unresolved sources.  The selection, however, was based on single-epoch photometry (and not on coadded images as  for the present study), making it less sensitive to morphology than the current one.  
The  morphology incompleteness $1/\epsilon_{\rm morph}$ is a factor  $\sim 1.7$ at redshift $z<0.5$, a factor 1.2 for $z$ in $0.5-0.8$,  1.1 for $z$ in $0.8-1.0$ and is compatible with 1.0 for $z>1$.  Note that we start our measurement of the quasar LF at $z=0.67$, where the correction is at most of order $\sim$10\%.

\subsubsection*{Target selection completeness, $\epsilon_{\rm sel} (g,\, m_{\rm diff})$:}  

\begin{figure}[htbp]
\begin{center}
\epsfig{figure= 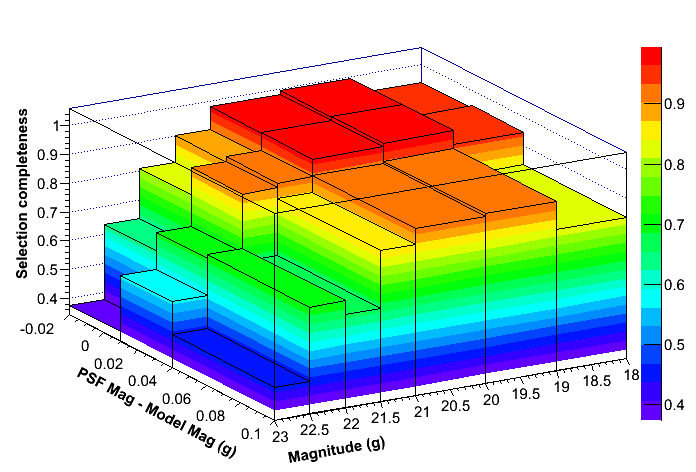,width = \linewidth} 
\caption{\it Selection completeness $\epsilon_{\rm sel} (g,\, m_{\rm diff})$ for quasars used in this study.  The loss at large PSF-model magnitudes, i.e., more extended objects,  is due to the more stringent cut on variability for such sources. The loss at large magnitudes is due to increased photometric dispersion in the lightcurves, which blurs the variability signal.  } 
\label{fig:effsel}
\end{center}
\end{figure}

The  selection completeness is determined from the fraction of  quasars in the control sample that pass the color and variability criteria of Sec.~\ref{subsec:ts}. The result is illustrated in Fig.~\ref{fig:effsel} as a function of magnitude $g$ and morphology $m_{\rm diff}$. The efficiency drops for fainter objects where the variability signal is not as visible, and for large $m_{\rm diff}$ because of the stricter variability cut applied to more extended objects. On average over the selected sample, the selection completeness is 0.86. 

\subsubsection*{Spectroscopic completeness, $\epsilon_{\rm spect}({g})$:}

Some spectra did not produce a reliable identification of the source, either because the extraction procedure  failed (yielding flat and useless spectra) or because the spectrum had too low a signal-to-noise ratio (S/N) for adequate identification, whether at the pipeline or at the visual inspection level.  As explained in Sec.~\ref{sec:eBOSS_var}, we globally consider such spectra  as inconclusive. We make the assumption that the ratio of quasars to non-quasars in the identified and  inconclusive sets are identical. This hypothesis is confirmed by the fact that the ensemble structure function parameters (globally accounted for by the $v_{\rm NN}$ parameter) are similar for the two sets. There is a small trend for a higher fraction of non-quasars in the inconclusive sample as the magnitude increases beyond $g\sim 22$, but the effect is small. Furthermore, because the selection completeness drops much faster with magnitude than the spectroscopic completeness (cf. Fig.~\ref{fig:completeness} for a comparative illustration), even a  large change in the fraction of quasars in the inconclusive set (for instance  from the current $ 80\%$ at $g\sim22.5$ to 50\%) only affects the overall completeness at $g\sim 22.5$ by less than 3\%. 

\begin{figure}[htbp]
\begin{center}
\epsfig{figure= 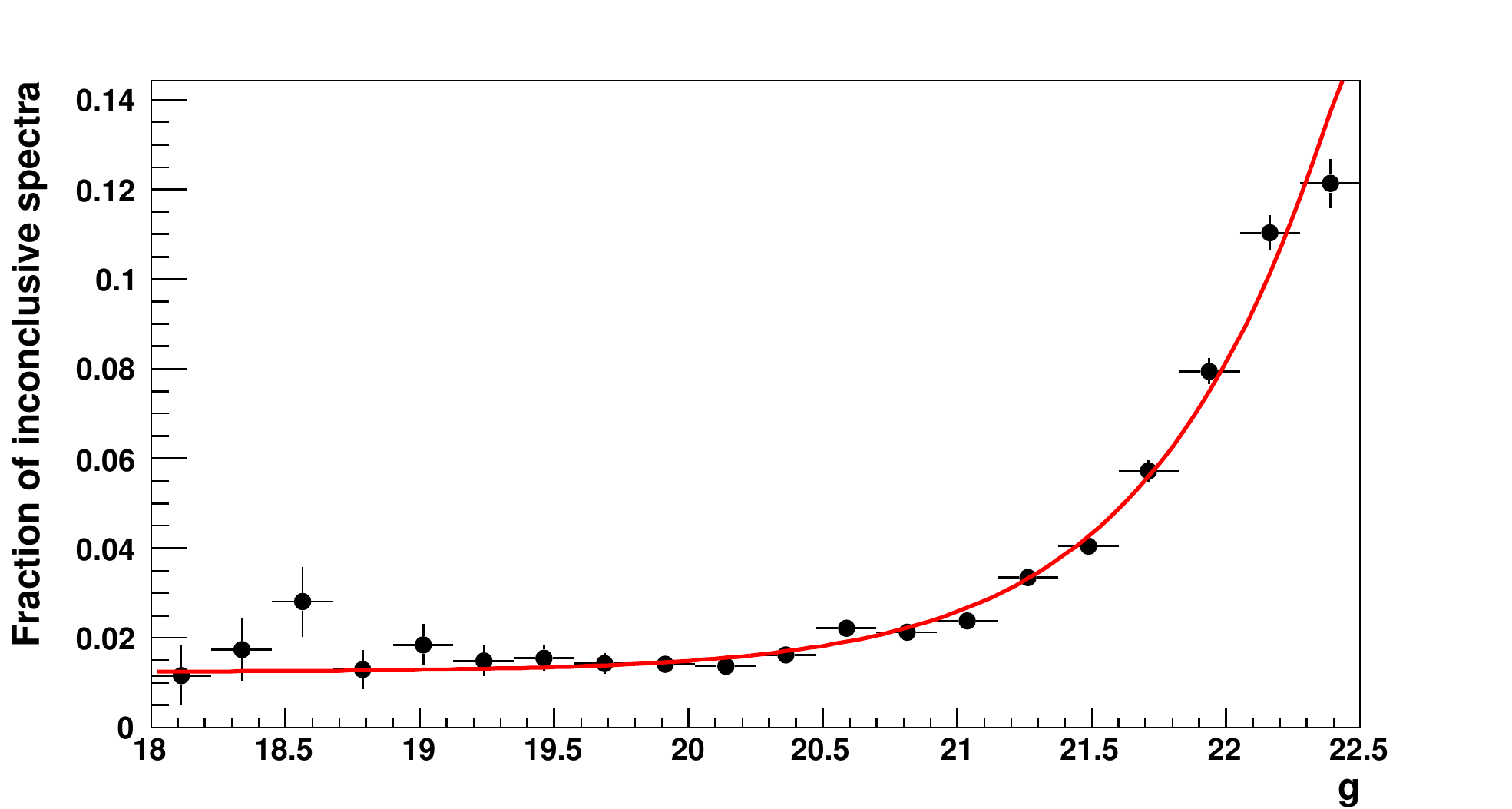,width = \linewidth} 
\caption{\it Fraction of inconclusive spectra (i.e., $1-\epsilon_{\rm spect}({g})$)  as a function of observed $g$ magnitude. The curve is an empirical fit to the data by a hyperbolic tangent. The $\sim 1\%$ loss at bright magnitudes is an artifact of the current pipeline. The increase at faint magnitudes is compatible with previous estimates.}
\label{fig:effspec}
\end{center}
\end{figure}

We compute the spectroscopic completeness from the fraction of inconclusive spectra as a function of magnitude (cf. Fig.~\ref{fig:effspec}).  This correction is only applied to new quasars since previously known ones are, by definition, spectroscopically  identified.  The completeness correction is 0.93 on average over the 7900 new quasars, and 1 by definition for known quasars. As illustrated in  Fig.~\ref{fig:effspec}, there is a constant $\sim$~1\% fraction of inconclusive spectra at bright magnitudes. This fraction increases to 8\% at $g=22$,  16\% at $g=22.5$ and 30\% at $g=23$. By comparison, the  measured fractions of inconclusive spectra are $3\%$, $8\%$ and $24\%$ at $g$ of 22, 22.5 and 23, respectively, in the BOSS+MMT program of Paper~LF. 
Part of the difference can be explained by the fact that we are now more conservative in the identification procedure. 
With identical definitions of inconclusive spectra, the percentages are compatible for $g>22.5$, but remain slightly larger for the current analysis at $g<22$. Inspection of the inconclusive spectra at bright magnitudes indicates that most of them are at redshift around 0.7 where the automated pipeline is not yet fully optimized (focus for BOSS was on $z>2.2$ quasars, and for eBOSS on $z>0.9$ quasars, thus higher redshifts than where this artefact appears). 

\subsubsection*{Tiling completeness, $\epsilon_{\rm tiling}$:}

Some spectroscopic targets cannot be observed either by lack of an available fiber on the plate (target density locally too high), or because a fiber cannot be placed at that plate position, e.g., because of fiber collisions~\citep[two fibers cannot be located less than 62$''$ apart, c.f. ][]{Dawson2013}. This loss is random, independent of redshift or magnitude, and $\epsilon_{\rm tiling}$ simply indicates the fraction of targets that were assigned fibers. It is equal to 0.959 for all new quasars, and equal to 1 by definition for already known quasars.

\subsection{Raw number counts}
\begin{figure}[htbp]
\begin{center}
\epsfig{figure= 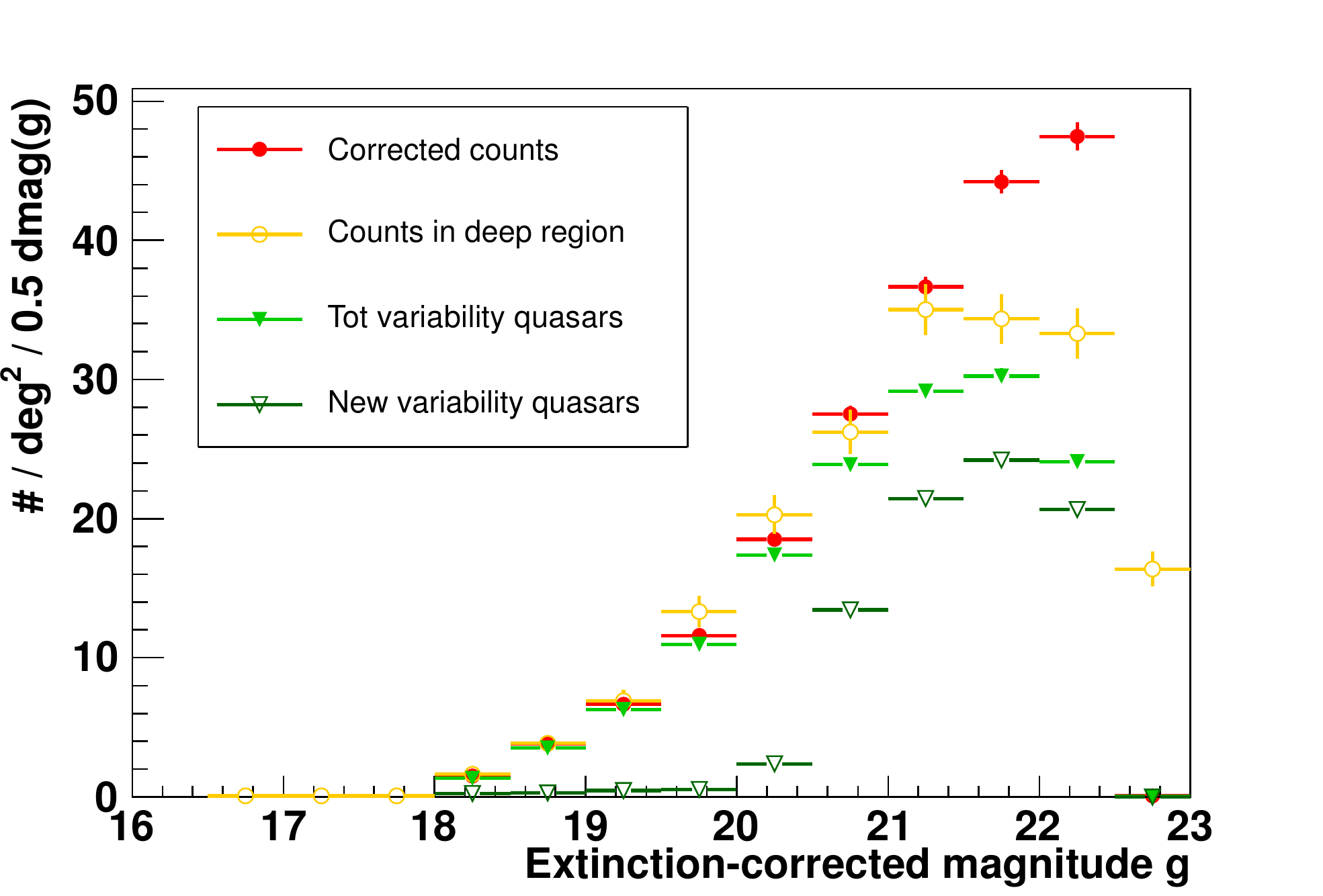,width = \linewidth} 
\epsfig{figure= 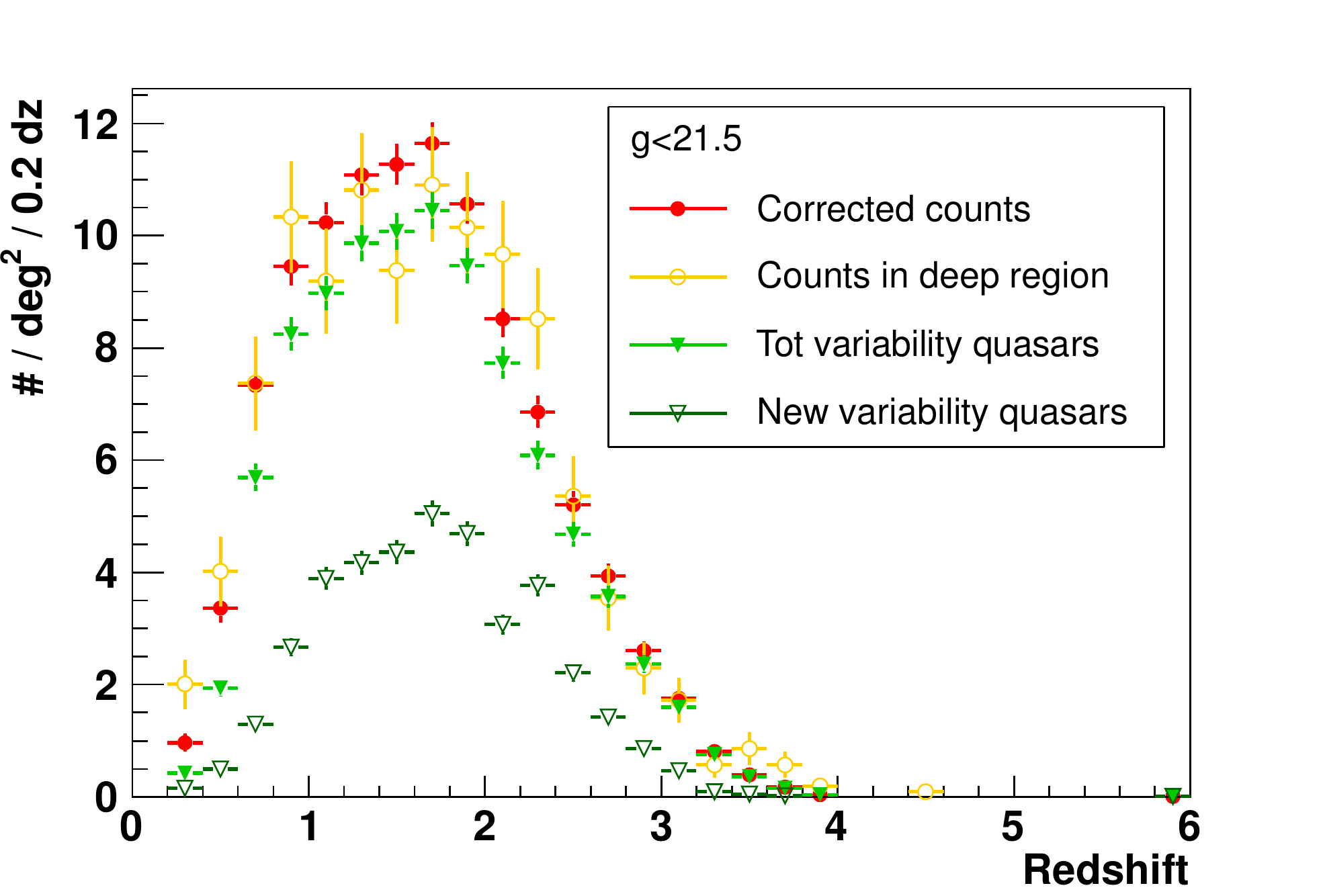,width = \linewidth} 
\epsfig{figure= 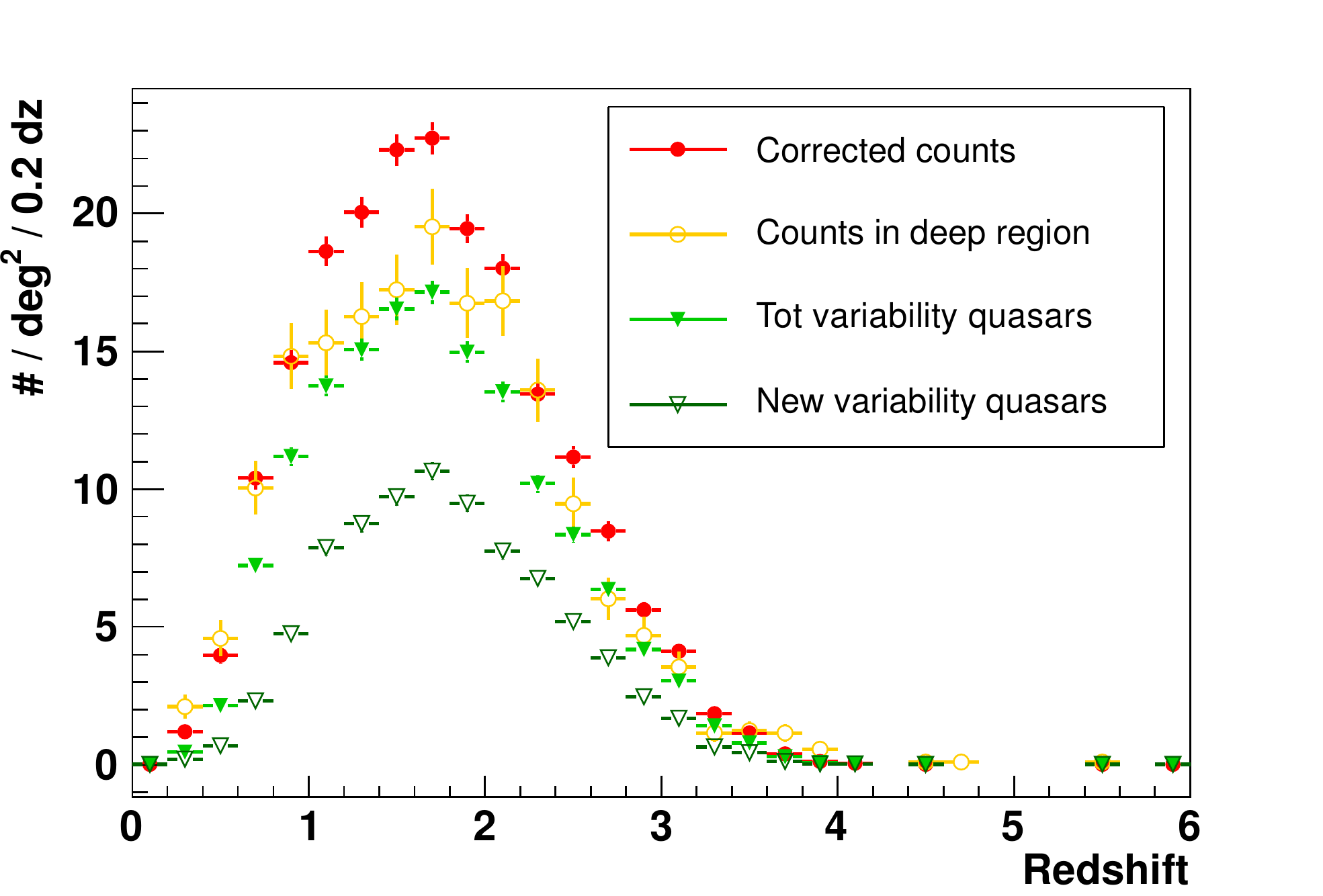,width = \linewidth} 
\caption{\it Extinction-corrected magnitude (top) and redshift (middle and bottom) distributions.
At faint magnitudes, most  variability-selected sample (green triangles) comes from the newly-identified quasars  (open dark green triangles). The deep sample (orange circles) from the $36^\circ<\alpha<42^\circ$ zone (cf. Sec.~\ref{subsec:data}) reproduces well the corrected counts (plain red circles) to $g=21.5$,  validating the computation of the completeness corrections to this magnitude limit. The deep sample is  the only one to extend beyond $g=22.5$. } 
\label{fig:dndg_dndz}
\end{center}
\end{figure}

We identified 7900 new quasars, and selected another 5976 that had  been previously spectroscopically identified. The magnitude and redshift distributions of the quasars selected by this study are shown in Fig.~\ref{fig:dndg_dndz} as the open dark green triangles for the new quasars, and as the plain green triangles for the total sample of selected quasars including the already identified ones. The deep sample (cf., Sec.~\ref{subsec:data}) is the only one to include quasars beyond $g_{\rm dered}=22.5$. 

As is clearly visible from Fig.~\ref{fig:dndg_dndz}, the newly identified quasars have a similar redshift distribution as the previously identified ones, but extend  to fainter magnitudes on average, with $\langle g_{\rm dered} \rangle=21.5~/~20.4$ for the newly / previously identified quasars, respectively.
Tab.~\ref{tab:rawcounts} lists the raw quasar counts for this study, in three bins of observed magnitude $g$. 
Our sample is particularly valuable at faint magnitudes: at  $g>22$, it  increases the number of  known  quasars in the footprint by almost a factor of 6.

\begin{table}[htb]
\begin{center}
\begin{tabular}{lcccc}
\hline\hline
&\multicolumn{3}{c}{Observed $g$ magnitude}&\\
Sample&$<21$&$21-22$&$>22$&Total\\
\hline\\[-8pt]
eBOSS & 1206 &4174&2520&7900\\
Known & 4155 & 1389& 432 & 5976\\
  \hline
  \end{tabular}
  \end{center}
\caption{Raw number counts for the samples of new (`eBOSS') and previously identified (`Known') quasars for the variability-selection of this study, in several $g$ magnitude bins.}
\label{tab:rawcounts}
\end{table}

\subsection{Corrected number counts} \label{sec:corrcounts}

\begin{figure}[htbp]
\begin{center}
\epsfig{figure= 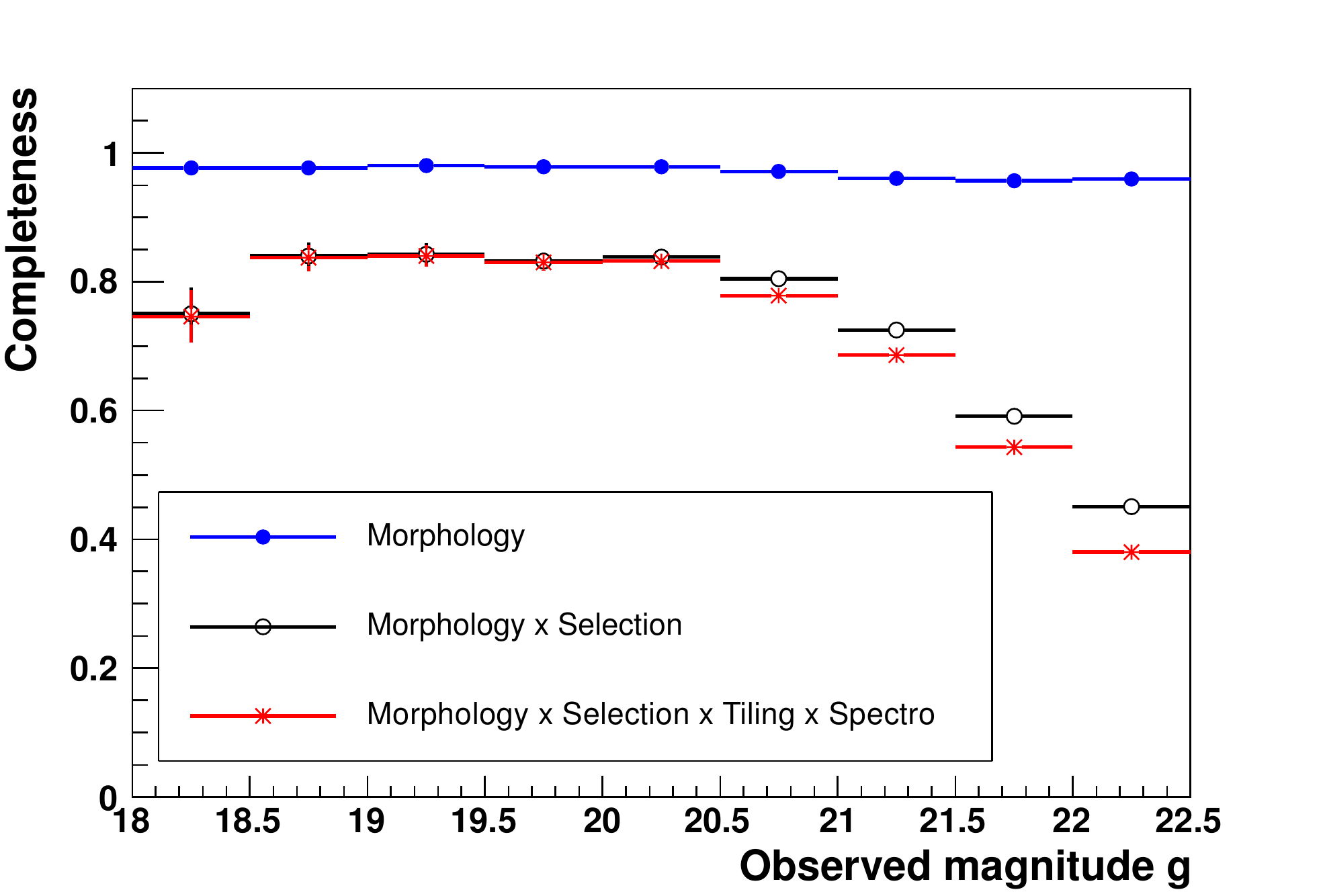,width = \linewidth} 
\caption{\it Magnitude dependence of the contributions to the total completeness correction, averaged over all variability-selected quasars of the eBOSS and known samples. } 
\label{fig:completeness}
\end{center}
\end{figure}

We derive corrected quasar number counts from  raw number counts by accounting for the different sources of incompleteness detailed above. New quasars are corrected for  morphology cut, target selection, tiling losses and spectroscopic failures, while previously known quasars are only corrected for morphology and  selection since their identification does not depend on eBOSS observation. 
The completeness-corrected number of quasars is thus given by
\begin{eqnarray}
N_{QSO} & = & \sum_{N_{\rm eBOSS}} \frac{1}{\epsilon_{\rm sel}(g,\, m_{\rm diff})\; \epsilon_{\rm morph}(z,g)\;\epsilon_{\rm tiling}\;\epsilon_{\rm spect}({g})} \nonumber \\
                &+&\sum_{N_{\rm Known}}\frac{1}{\epsilon_{\rm  sel}(g,\, m_{\rm diff})\; \epsilon_{\rm  morph}(z,g)} ~.
\end{eqnarray}
The total completeness correction for new eBOSS quasars  has an average of 0.70 and a standard deviation of 0.15, and for previously known quasars an average of 0.90 and a standard deviation of 0.12. The overall completeness of the full sample as a function of magnitude is illustrated in Fig.~\ref{fig:completeness}.

To validate the  computation of the completeness corrections, it is interesting to compare the corrected quasar densities to the    measurements done in  the deep zone. Despite larger counts, it is clearly visible in Fig.~\ref{fig:dndg_dndz} (top plot) that the deep region still suffers from significant incompleteness at $g_{\rm dered}>21.5$. In Tab.~\ref{tab:counts}, we therefore provide both the completeness-corrected densities to a limiting magnitude $g_{\rm dered}<21.5$ where the deep sample is expected to be complete, and those to $g_{\rm dered}<22.5$ from which we  measure the quasar LF. At $g_{\rm dered}<21.5$, the corrected counts and the deep sample show excellent consistency, with less than 1.5~$\sigma$ deviations  over the entire redshift range (Fig.~\ref{fig:dndg_dndz}, middle plot), compatible with Poisson errors. 
 The magnitude distributions  of the completeness-corrected and deep-zone counts   (Fig.~\ref{fig:dndg_dndz}, top plot)  are also in excellent agreement  to $g_{\rm dered}<21.5$. Extending the study to $g_{\rm dered}<22.5$ (Tab.~\ref{tab:counts} or Fig.~\ref{fig:dndg_dndz}, bottom plot), the corrected counts are of order 10\% larger  than the counts in the deep zone, a reasonable excess that is easily accounted for by the incompleteness of the deep sample at the faint end.

\begin{table}[htb]
\begin{center}
\begin{tabular}{lcccc}
\hline\hline
&\multicolumn{3}{c}{Redshift range}&\\
Source&$z<0.9$&$0.9<z<2.1$&$z>2.1$&Total\\
\hline\\[-8pt]
$g_{\rm dered}<21.5$&&&\\
This work& 16 $\pm 1$ & 64 $\pm 1$ & 26 $\pm 1$ & 106 $\pm 1$ \\
Deep zone & $18\pm 1$ & 60 $\pm 2$ & 29 $\pm 2$ & 108 $\pm 3$ \\
Paper LF & 16  & 50 & 21 & 87 \\
 \hline\\[-8pt]
$g_{\rm dered}<22.5$&&&\\
This work& 23 $\pm 1$ & 119 $\pm 1$ & 56 $\pm 1$ & 198 $\pm 1$ \\
Deep zone & $23\pm 1$ & 101 $\pm 3$ & 51 $\pm 2$ & 175 $\pm 4$ \\
Paper LF & 27  & 96 & 48 & 171 \\
 \hline
  \end{tabular}
  \end{center}
\caption{Areal densities (in $\rm deg^{-2}$) in several redshift bins for this work (after completeness correction), for the deep $36^\circ<\alpha<42^\circ$ zone (raw counts) and from the best-fit luminosity function of Paper~LF.
Uncertainties on  raw counts are  Poisson noise.}
\label{tab:counts}
\end{table}

Tab.~\ref{tab:counts} also lists the expected number counts from the luminosity function computed in Paper~LF. The results from this new study show a small global  increase of 10 to 20\% in number counts over these previous estimates. As explained above, the completeness-corrected densities measured in this work  are in agreement with the deep-zone raw counts  at the bright end, and  show a 10\% understandable excess at the faint end. 
Furthermore, the deep sample gives a solid lower bound on quasar densities. 
For these reasons, we are confident that the discrepancy between the two studies is due to slightly underestimated completeness corrections in Paper~LF, rather than overestimated corrections in the present work. Given that the total corrections in Paper~LF ranged from 1.2 at $g<20$ to a factor 2 at $g\sim 22.5$, a 10\% inaccuracy is not surprising.


\section{Luminosity function in g}
\label{sec:LF}

We derive the QLF in a  similar manner as described in Paper~LF. We compute a binned QLF from the corrected number counts of Sec.~\ref{sec:corrcounts}, considering our completeness limit at $g_{\rm dered}=22.5$.  We  fit two  parametric models to our binned QLF and compare the  number counts each model predicts over the  range of magnitude and redshift observable by eBOSS. Finally, we use our QLF fits to predict number counts to fainter magnitudes than achieved by eBOSS, as needed for future quasar surveys.

\subsection{Binned luminosity function}

\begin{figure}[htbp]
\begin{center}
\epsfig{figure= 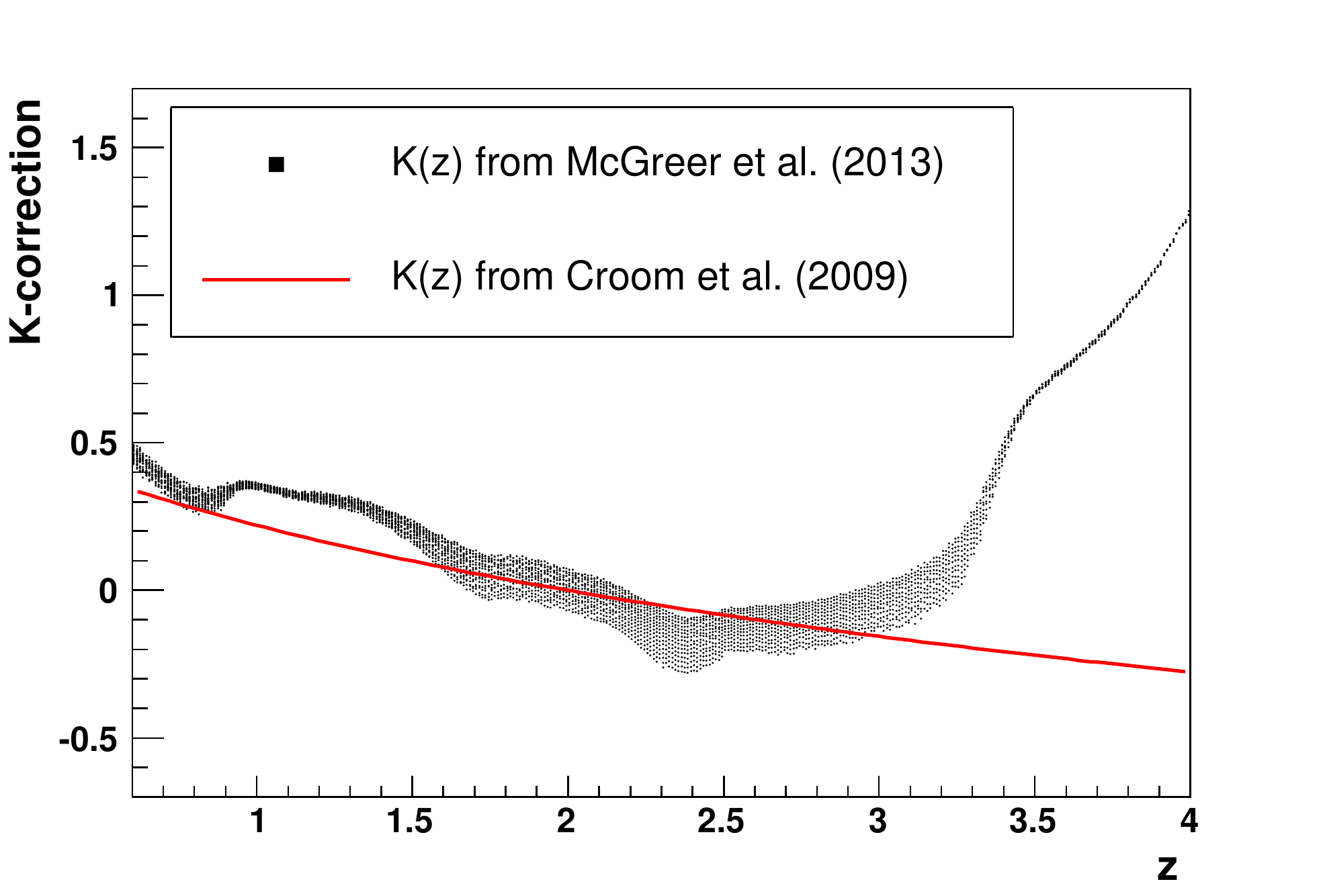,width = \linewidth} 
\caption{\it The $g$-band K-correction   as a function of redshift. The spread illustrates the luminosity variation of the correction for the quasars in our sample. The sudden change 	at $z>3$ is due to the suppression of the flux in $g$-band by the intervening Lyman-$\alpha$ absorbers.   } 
\label{fig:Kcorr}
\end{center}
\end{figure}

Selection for this survey was performed in the $g$-band, which provides the highest S/N for a vast fraction of the data. For each  quasar, we compute the absolute magnitude normalized to $z=2$ by
\begin {equation}
M_g(z\!=\!2) = g_{\rm dered}-d_M(z)-[K(z)-K(z\!=\!2)]~,
\end{equation}
where the distance modulus $d_M(z)$ is computed assuming a  flat $\Lambda$CDM cosmology with $(\Omega_\Lambda,\Omega_M,w,h) = (0.6935,0.3065,-1,0.679)$ as measured by the~\citet{Planck2015} in the `TT+lowP+lensing+ext' configuration, and $K(z)$ is the $K$-correction that accounts for redshifting of the bandpass of the spectrum. We choose to normalize the magnitudes at $z=2$ because it is close to the median redshift of our sample, and  it allows backward compatibility  with previous studies~\citep{Croom2009, Ross2013,Palanque2013a}. As in Paper~LF, we use the $K$-correction as a function  of redshift that was derived by ~\citet{McGreer2013} following a similar approach as in \cite{Richards2009}. As  illustrated in Fig.~\ref{fig:Kcorr}, the $K$-correction is  close to that of \citet{Croom2009} for $z<3$, but extends to higher redshifts. It varies with luminosity, due to the accounting of strong quasar emission lines whose equivalent widths are a function of luminosity~\citep{Baldwin1977}. This luminosity dependence introduces a spread of $\sim 0.25$ mag at $z\sim 2-3$ where the Lyman-$\alpha$ and  C-IV lines contribute substantially to the flux in $g$-band.

We define eight redshift bins, with limits 0.68, 1.06, 1.44, 1.82, 2.2, 2.6, 3.0, 3.5 and 4.0.  The binned LF is computed  for each redshift using the model-weighted estimator $\Phi$ suggested by \citet{Miyaji2001} and used in previous studies such as in \citet{Croom2009} or in Paper~LF. The binned LF is given by
\begin{equation}
\Phi(M_{g_i},z_i)\;=\;\Phi^{model}(M_{g_i},z_i)\;\frac{N_i^{obs}}{N_i^{model}}\; ,
\label{eq:phimodel}
\end{equation}
where $M_{g_i}$ and $z_i$ are, respectively, the absolute magnitude and the redshift at the center of bin $i$,
$\Phi^{model}$ is the model LF estimated at the center of the bin,
$N_i^{model}$ is the number of quasars in the bin with $g_{\rm dered}<22.5$ estimated from the model,  and $N_i^{obs}$ is the observed number of quasars in the bin. 
This estimator is free from most of the biases unavoidable in the more usual $1/V$ method  devised by \citet{Page2000}, and  improves upon it in several ways. It corrects for variations of the LF within a bin (particularly critical at the steep bright end of the QLF), corrects for incompleteness in a bin (particularly critical at the faint end of our LF where the bin is incompletely sampled), and allows exact errors to be evaluated using Poisson statistics.   
A drawback of this estimator is that it is model-dependent, but \citet{Miyaji2001} demonstrated that the uncertainties  due to the model dependence are practically negligible.  We here free ourselves of any model-dependence by performing an iterative fitting to determine the binned QLF from Eq.~\ref{eq:phimodel} and our choice of model, until parameter convergence is reached. As explained in the following section, we use two different models to fit the QLF, which do not produce any significant difference to the estimate of the binned LF.


\subsection{QLF model fits}
\label{sec:QLF}

The QLF is traditionally fit by a double power law of the form~\citep{Boyle2000, Richards2006}:
\begin{equation}
\Phi(M_g,z)\;=\;\frac{\Phi^*}{10^{0.4(\alpha+1)(M_g-M_g^*)}\;+\;10^{0.4(\beta+1)(M_g-M_g^*)}}
\label{eq:phi}
\end{equation}
where $\Phi$ is the quasar comoving space density and $M_g^*$ a characteristic, or break, magnitude. The slopes $\alpha$ and $\beta$ describe the evolution of the LF on either side of the break magnitude. In this work, we consider two extensions of this simple form  described below. 

Our first model is the same as the one we used in Paper~LF. We  consider a pure luminosity-evolution (PLE) model as in~\citet{Croom2009}, where a redshift dependence of the luminosity is introduced through an evolution in  $M_g^*$  described by 
\begin{equation}
M_g^*(z)=M_g^*(z_p)-2.5[k_1(z-z_p)+k_2 (z-z_p)^2]\; .
\label{eq:mg}
\end{equation}
We allow the redshift-evolution parameters ($k_1$ and $k_2$) and the model slopes ($\alpha$ and $\beta$) to be different on either side of a pivot redshift $z_p=2.2$. The model is thus  described by Eqs.~\ref{eq:phi} and \ref{eq:mg} where $\alpha$, $\beta$, $k_1$ and $k_2$  are defined with subscript $l$ for $z<z_p$ and $h$ for $z>z_p$. This PLE model therefore has ten parameters ($\Phi^*$, $M_g^*(z_p)$, $\alpha_l$, $\beta_l$, $k_{1l}$, $k_{2l}$, $\alpha_h$, $\beta_h$, $k_{1h}$ and $k_{2h}$) that are  free to vary in the fit.

An extensive study of the QLF was performed by \citet{Ross2013} on  23,300 quasars with $i<21.8$ and $2.2<z<3.5$ from the DR9 release of BOSS data~\citep{Ahn2012}, complemented by about  $ 5500$  quasars over $2.2<z<3.5$ from Paper~Var and about $ 1900$ quasars over $0.3<z<3.5$  from Paper~LF.  The authors showed that a good fit to this large sample of quasars over the full redshift range was obtained by using a PLE model for $z<2.2$,
and a model with both luminosity and density evolution (LEDE) for $z>2.2$, where the normalization and break magnitude evolve in a log-linear manner, e.g., 
\begin{eqnarray}
\log[\Phi^*(z)]&=&\log[\Phi^*(z_p)]+c_{1a}(z-z_p)+c_{1b}(z-z_p)^2\\
M_g^*(z)&=&M_g^*(z_p)+c_2(z-z_p)
\label{eq:mg_LEDE}
\end{eqnarray} 
with $z_p=2.2$ the pivot redshift. 
This PLE (for $z<2.2$) + LEDE (for $z>2.2$) is our second model with which we fit our binned QLF. Unlike  \citet{Ross2013}, however, we impose continuity of the LF at $z=z_p$ by requiring the same normalization $\Phi^*(z_p)$ and break magnitude $M_g^*(z_p)$ for both the PLE and the LEDE forms. We allow for some additional flexibility by allowing a redshift dependence of the slope, according to
\begin{equation}
\alpha(z) = \alpha(z_p) + c_3(z-z_p)\;,
\end{equation}
where $\alpha(z_p)$ is equal to the value of $\alpha$ used at $z<z_p$ in the PLE form.
Our PLE+LEDE model therefore has ten parameters ($\Phi^*(0)$, $M_g^*(0)$, $\alpha (z_p)$, $\beta (z_p)$, $k_1$, $k_2$, $c_{1a}$, $c_{1b}$, $c_2$ and $c_3$) that are left free to vary in the fit. 

\begin{table*}[htb]
\begin{center}
\begin{tabular}{ccccccccc}
\hline\hline
\multirow{2}{*}{Model}&Redshift &\multicolumn{6}{c}{ \multirow{2}{*}{Parameters} } & \multirow{2}{*}{$\chi^2/\nu$}\\
& range\\
\hline\\[-8pt]

\multirow{5}{*}{PLE}& &$M_g^*(z_p)$& $\log (\Phi^*)$& \\
& $0.68-4.0$ &$-26.71{\scriptstyle \pm0.15}$ & $-6.01{\scriptstyle \pm0.07}$ & & & & & 135/76\\

& & & & $\alpha$& $\beta$&$k_1$&$k_2$ \\
&$0.68-2.2$ &  & & $-4.31{\scriptstyle \pm0.26}$&$-1.54{\scriptstyle \pm0.04}$ & $-0.08{\scriptstyle \pm 0.08}$ & $-0.40{\scriptstyle \pm 0.05}$\\

 & $2.2-4.0$ & & & $-3.04{\scriptstyle \pm 0.12}$ & $-1.38{\scriptstyle \pm 0.07}$ & $-0.25{\scriptstyle \pm 0.09}$ &$-0.05{\scriptstyle \pm 0.06}$ \\
 
 \hline\\[-8pt]
 
&&  $M_g^*(0)$ & $\log [\Phi^*(0)]$& $\alpha$& $\beta$ \\

& $0.68-4.0$  & $-22.25{\scriptstyle \pm0.49}$ &$-5.93{\scriptstyle \pm0.09}$& $-3.89{\scriptstyle \pm 0.23}$ & $-1.47 {\scriptstyle \pm0.06}$  & &  &146/77\\
  
PLE& & $k_1$ & $k_2$ \\
+ LEDE& $0.68-2.2$ & $1.59 {\scriptstyle \pm 0.28}$ & $-0.36 {\scriptstyle \pm 0.09}$ \\

& & $c_{1a}$ & $c_{1b}$ & $c_2$ & $c_3$ \\
& $2.2-4.0$ & $-0.46 {\scriptstyle \pm 0.10}$ &$ -0.06 {\scriptstyle \pm 0.10}$ & $-0.14 {\scriptstyle \pm  0.17}$ & $0.32 {\scriptstyle \pm 0.23}$ \\
\hline
  \end{tabular}
  \end{center}
\caption{Values of the parameters (and redshift range over which they apply) for the best-fit PLE and PLE+LEDE models of quasar luminosity functions (e.g., Eqs.~\ref{eq:phi}--\ref{eq:mg_LEDE}). The slope $\alpha$ reproduces the bright end part of the QLF, and $\beta$ the faint end. }
\label{tab:fits}
\end{table*}

\begin{figure*}[htbp]
\begin{center}
\epsfig{figure=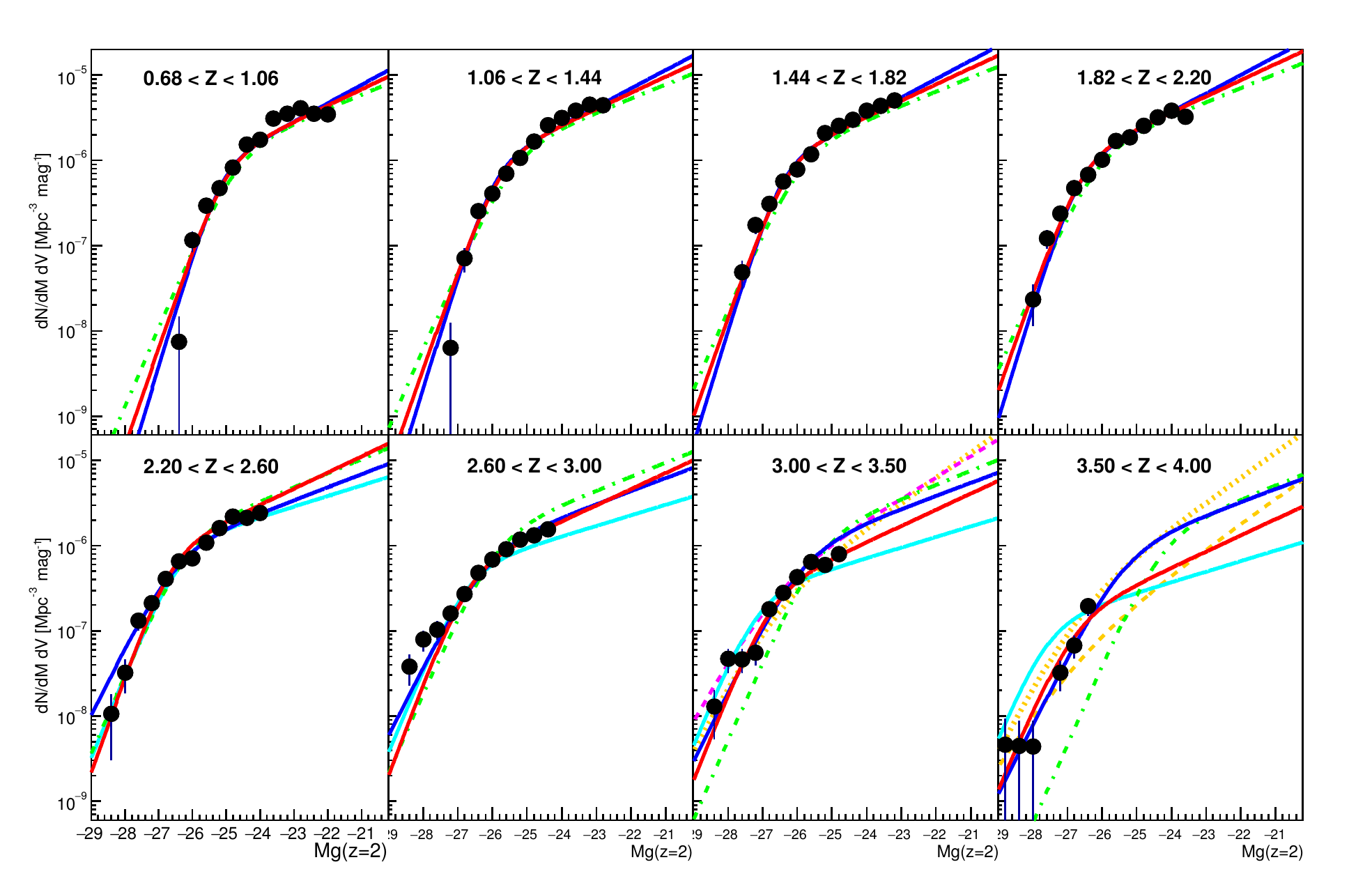,width = \textwidth} 
\caption[]{
Quasar luminosity function measurements (black circles). The best-fit models of this work are shown as the red (respectively blue) curves for the PLE+LEDE (resp. PLE) models. The green dot-dashed curve is the LF of \citet{Croom2009}. The plain cyan curve is the best fit LEDE model of \cite{Ross2013} at $z>2.2$. 
The orange dotted and dashed curves are the best fits to COSMOS data~\citep{Masters2012} at $z\sim 3.2$ (shown in the last two redshift bins) and $z\sim 4.0$, respectively. 
The  magenta dashed curve (almost exactly overlapping the orange dotted curve at the faint end in the $3.0<z<3.5$ redshift bin) is measured at $z\sim 3.2$ from SWIRE and SDSS \citep{Siana2008}. 
} 
\label{fig:clum}
\end{center}
\end{figure*}

\begin{figure}[htbp]
\begin{center}
\epsfig{figure=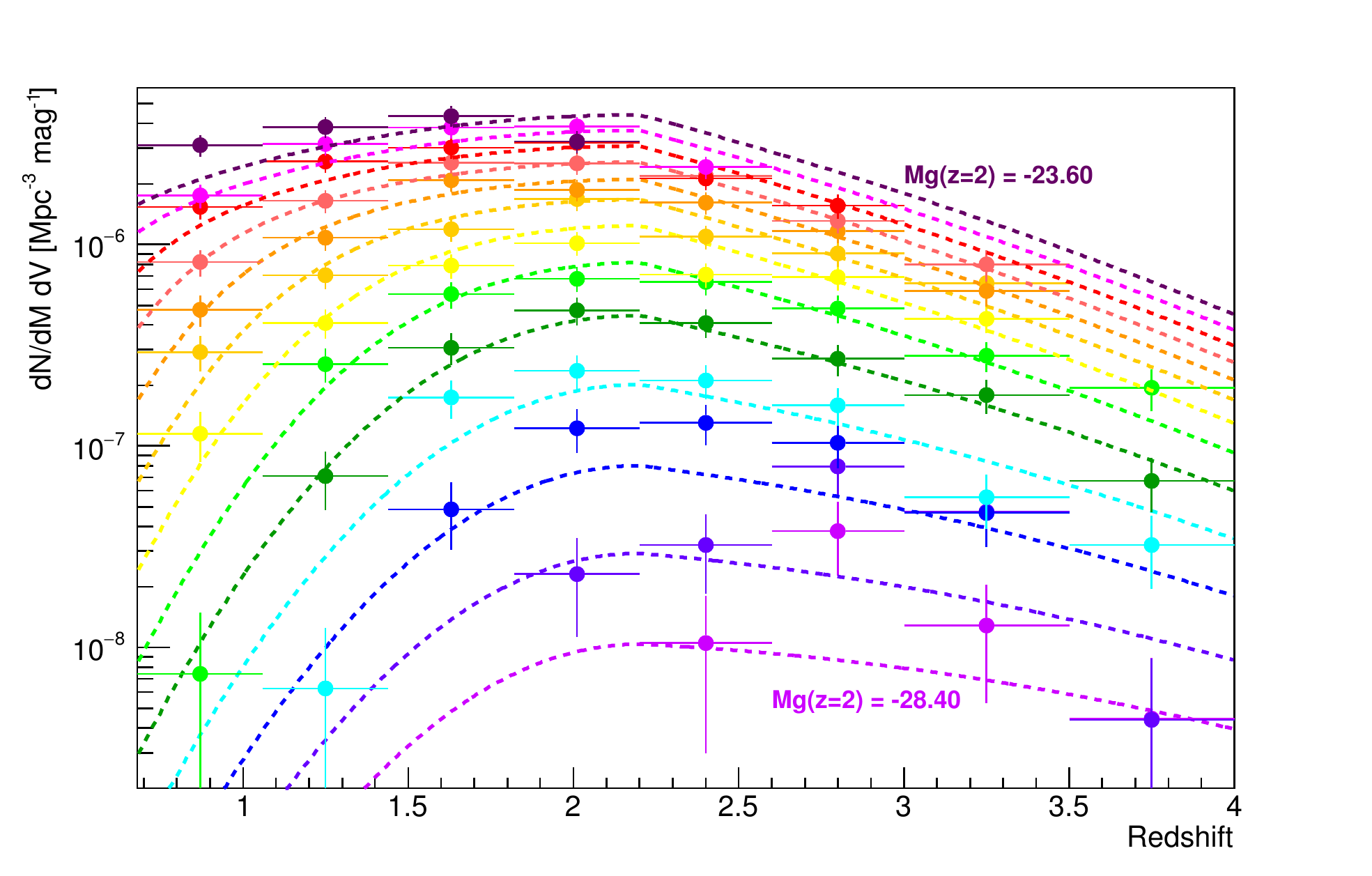,width = \columnwidth} 
\caption[]{. Comparison of our best-fit PLE+LEDE model (solid curves) to the eBOSS Stripe 82 QLF data (points). From bottom to top, the curves correspond to magnitudes $M_G(z\!=\!2) = -28.4$, $-28$, $-27.6$, $-27.2$, $-26.8$, $-26.4$, $-26$, $-25.6$, $-25.2$, $-24.8$, $-24.2$, $-24$ and $-23.6$. A simple PLE model  breaks down at $z=2.2$. Here, we switch  to an LEDE model at $z>2.2$. The mixed PLE+LEDE model reproduces the QLF well.
} 
\label{fig:fctlumi_ROSS4}
\end{center}
\end{figure}

The best-fit parameters are given in Tab.~\ref{tab:fits}. Both models have ten parameters free to vary in the fit, and start with 100 data points spread over eight redshift bins. Throughout the iterations (a total of ten iterations is needed in both cases to reach parameter convergence), the QLF is recomputed according to Eq.~\ref{eq:phimodel}, and points that are corrected by more than a factor 2 (due to the incompleteness introduced by the $g_{\rm dered}<22.5$ cut) are removed from the fit. One to two points per redshift bin are  excluded by this procedure. The resulting best-fit models are illustrated in Fig.~\ref{fig:clum}, along with the best-fit model obtained by \citet{Croom2009} for $z<2.2$ and extrapolated to higher redshift, and the PLE+LEDE model of \citet{Ross2013} shown for $z>2.2$, where it is  best constrained by the DR9 data  used for the fit. In the $z<2.2$ range, our two models are in excellent agreement with \citet{Croom2009}, and are a good fit to our binned QLF. At $z>2.2$, our two models show similar trends at the bright end, but start to differ at the faint end of the QLF, in particular for the highest two redshift bins, which are lacking faint quasars. The agreement with \citet{Ross2013} is good over the common redshift range, but the fits also start to deviate at $z>3.5$ where data are scarce. Although we constrain our models to be continuous at $z=2.2$, the fit reduced $\chi^2$s are less than 2. Such values were only obtained in \citet{Ross2013} when fitting over restricted redshift ranges, typically limiting to data below or above a redshift of 2.2. Fig.~\ref{fig:fctlumi_ROSS4} presents the redshift evolution of the QLF in a series of luminosity bins, including both our data and the best-fit PLE+LEDE model. 
The `kink' at $z=2.2$ is due to the change of analytical form at this pivot redshift. The model, however, is continuous at $z=2.2$.

We provide, in Appendix A, the measured luminosity function values and associated uncertainties, where the measurements were corrected using the model-weighted estimator of Eq.~\ref{eq:phimodel} and considering the best-fit PLE+LEDE  model.

\subsection{Predicted number counts}

\begin{figure}[htbp]
\begin{center}
\epsfig{figure=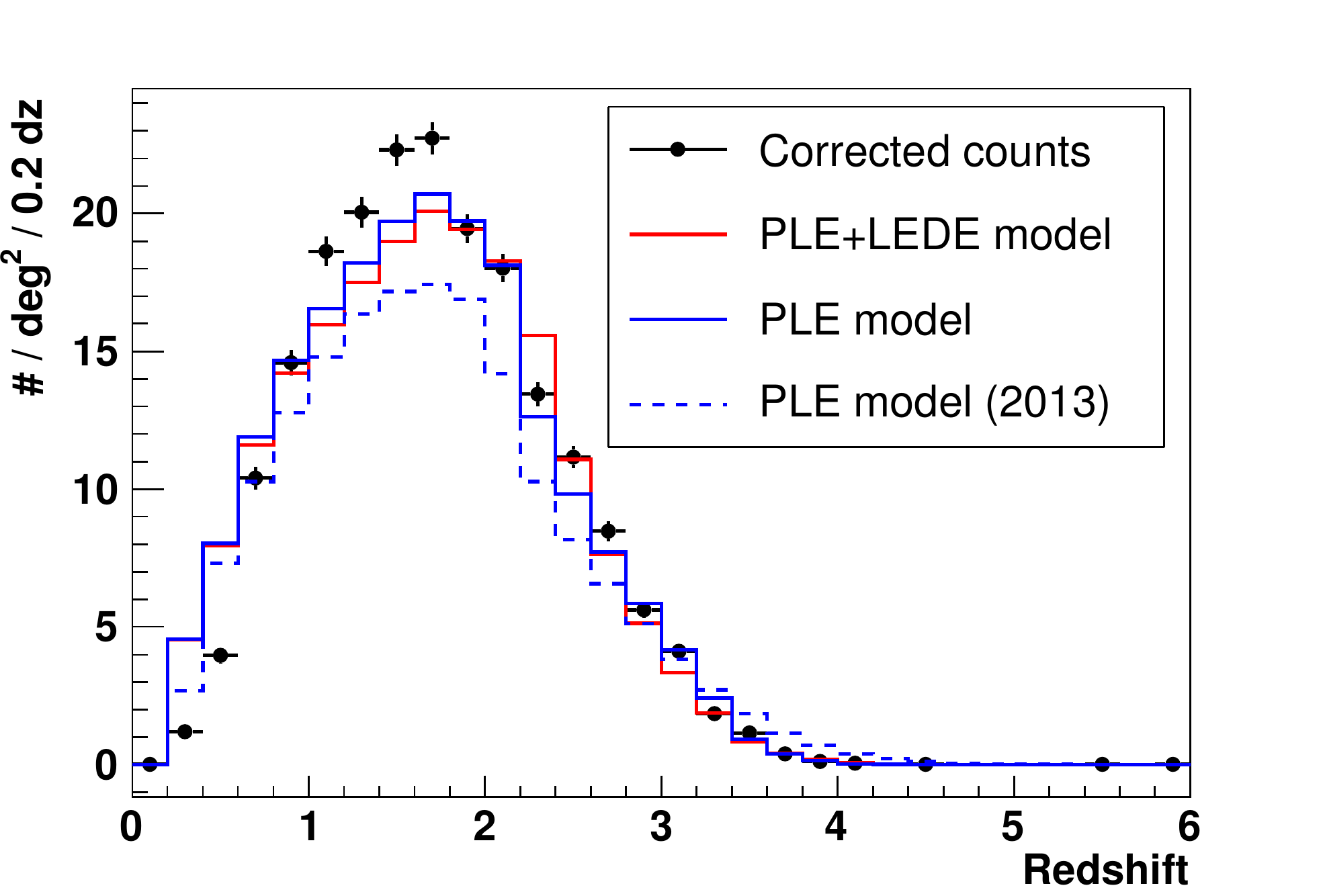,width = \columnwidth} 
\caption[]{Projected counts for $g_c<22.5$  for the PLE (blue) and the PLE+LEDE (red) luminosity function models used to fit the binned QLF. The dashed curve is the best-fit model of Paper~LF, shown here for comparison. Our new fits indicate a $\sim $~10\% increase in total  quasar counts.
} 
\label{fig:dndz_LF_g}
\end{center}
\end{figure}

Out best-fit luminosity functions can be used to estimate the density of quasars as a function of redshift and  magnitude. As shown in Tab.~\ref{tab:countsDESI},  the redshift distributions predicted by our two models agree  within 0.3\% for counts to a limiting magnitude  $g<22.5$, and within 1.1\% to $g<23$. The largest discrepancy between the models occurs near $z=2.3$ (cf. Fig.~\ref{fig:dndz_LF_g}), close to the pivot redshift $z_p=2.2$ where the analytical form of our models changes, causing a  small discontinuity in the derivative of their redshift evolution. At $z\sim 2.3$,  the predictions from the two models differ by about 10\%.

The two models are also in good agreement 
with the corrected number counts of Sec.~\ref{sec:corrcounts} (Tab.~\ref{tab:counts}), and indicate a $\sim 15\%$ increase over previous estimates based on the QLF fit of Paper~LF.  Both the PLE and the PLE+LEDE models fit the $z>2$ data well, but do not provide as good a fit to the $z<0.6$ range. A possible explanation for this discrepancy is a loss of low-$z$ quasars in our selection, possibly due to the morphology cut for which the completeness efficiency $\epsilon_{\rm morph}$ does not sufficiently account. Another possibility is the misidentification of low-$z$ quasars with the current  pipeline (cf.  paragraph on spectroscopic completeness in Sec.~\ref{sec:completenesscor}). This small discrepancy, however, is of little relevance for large-scale spectroscopic  surveys that are mostly focusing on quasars at $z>1$ where they are abundant enough to probe matter clustering.

\begin{table}[htbp]
\begin{center}
\begin{tabular}{cccccc}
\hline\hline\\[-8pt]
\multirow{2}{*}{Model}  & \multirow{2}{*}{Limit} & \multicolumn{3}{c}{Redshift range} &\multirow{2}{*}{Total}\\
&& $<0.9$ & $0.9-2.1$&$>2.1$ &  \\  \hline\\[-8pt]
PLE & $g<22.5$ & 32 & 111 & 53 & 196\\
PLE+LEDE & $g<22.5$& 31 & 108 & 55 & 195\\ \hline\\[-8pt]

PLE & $g<23$ & 43 & 151 & 75 & 269\\
PLE+LEDE & $g<23$& 41& 145& 77& 263\\ \hline\\[-8pt]

PLE& $r<23$ & 49 & 171 & 91 & 311\\
PLE+LEDE  &$r<23$&46 & 162 & 88 & 296\\
\hline
 \end{tabular}
  \end{center}
\caption{Quasar densities (in $\rm deg^{-2}$) predicted from the two phenomenological models of this study. The PLE and the PLE+LEDE models give very similar density estimates.  }
\label{tab:countsDESI}
\end{table}

Using the average  $g-r$ vs. redshift dependence that we measure for all DR12Q BOSS quasars, we
can also calculate  the number of quasars as a function of $r$-band magnitude. We provide these  estimates for our two models as a function of $g$   in Tab.~\ref{tab:nb_g}, and as a function of $r$ in Tab.~\ref{tab:nb_r},  for an hypothetical survey covering 10,000~$\rm deg^2$. 

\begin{table*}[htbp]
\begin{center}
\begin{tabular}{crrrrrrr}
\hline\hline\\[-8pt]
\multicolumn{8}{c}{PLE model}\\
$g\setminus z$ & 0.5 & 1.5 & 2.5 & 3.5 & 4.5 & 5.5 & Total \\  \hline\\[-8pt]
15.75 & 26 & 2 & 10 & 0 & 0 & 0 & 38 \\ 
16.25 & 58 & 8 & 33 & 0 & 0 & 0 & 100 \\ 
16.75 & 143 & 39 & 87 & 3 & 0 & 0 & 272 \\ 
17.25 & 386 & 180 & 231 & 10 & 0 & 0 & 806 \\ 
17.75 & 1134 & 813 & 626 & 26 & 0 & 0 & 2599 \\ 
18.25 & 3396 & 3468 & 1727 & 68 & 0 & 0 & 8660 \\ 
18.75 & 8930 & 12525 & 4780 & 179 & 0 & 0 & 26413 \\ 
19.25 & 17963 & 32977 & 12319 & 462 & 1 & 0 & 63722 \\ 
19.75 & 28172 & 61064 & 26357 & 1170 & 3 & 0 & 116766 \\ 
20.25 & 38732 & 90121 & 45742 & 2827 & 7 & 0 & 177430 \\ 
20.75 & 50756 & 121102 & 68898 & 6299 & 18 & 0 & 247074 \\ 
21.25 & 65550 & 157999 & 95351 & 12540 & 47 & 0 & 331487 \\ 
21.75 & 84300 & 204485 & 125298 & 22098 & 120 & 0 & 436302 \\ 
22.25 & 108276 & 264163 & 159894 & 35046 & 301 & 1 & 567680 \\ 
22.75 & 138990 & 341219 & 201144 & 51351 & 737 & 3 & 733442 \\ 
23.25 & 178321 & 440913 & 251597 & 70978 & 1722 & 7 & 943538 \\ 
23.75 & 228612 & 570015 & 314273 & 93793 & 3739 & 17 & 1210449 \\ 
24.25 & 292750 & 737314 & 392821 & 119723 & 7357 & 42 & 1550006 \\ 
24.75 & 374172 & 954240 & 491772 & 149132 & 13010 & 106 & 1982433 \\ 
\hline\\[-8pt]
Total & 1620667 & 3992648 & 2192958 & 565706 & 27062 & 175 & 8399216\\ 
\hline\hline\\[-8pt]
\multicolumn{8}{c}{PLE+LEDE model}\\
$g\setminus z$ & 0.5 & 1.5 & 2.5 & 3.5 & 4.5 & 5.5 & Total \\  \hline\\[-8pt]
15.75 & 35 & 5 & 1 & 0 & 0 & 0 & 42 \\ 
16.25 & 82 & 20 & 5 & 0 & 0 & 0 & 108 \\ 
16.75 & 200 & 77 & 20 & 1 & 0 & 0 & 298 \\ 
17.25 & 519 & 291 & 76 & 3 & 0 & 0 & 889 \\ 
17.75 & 1402 & 1088 & 287 & 10 & 0 & 0 & 2787 \\ 
18.25 & 3758 & 3887 & 1065 & 35 & 0 & 0 & 8745 \\ 
18.75 & 9092 & 12345 & 3737 & 117 & 0 & 0 & 25291 \\ 
19.25 & 17972 & 31289 & 11415 & 382 & 1 & 0 & 61059 \\ 
19.75 & 28629 & 59594 & 27445 & 1163 & 4 & 0 & 116836 \\ 
20.25 & 39554 & 90405 & 50511 & 3077 & 13 & 0 & 183560 \\ 
20.75 & 51163 & 121609 & 76197 & 6648 & 36 & 1 & 255653 \\ 
21.25 & 64559 & 155923 & 103193 & 11780 & 97 & 2 & 335554 \\ 
21.75 & 80771 & 196666 & 133441 & 18077 & 242 & 4 & 429201 \\ 
22.25 & 100758 & 246795 & 169698 & 25349 & 543 & 9 & 543153 \\ 
22.75 & 125546 & 309293 & 214695 & 33716 & 1063 & 21 & 684334 \\ 
23.25 & 156321 & 387578 & 271321 & 43594 & 1820 & 46 & 860680 \\ 
23.75 & 194486 & 485819 & 342970 & 55625 & 2799 & 93 & 1081792 \\ 
24.25 & 241694 & 609219 & 433845 & 70598 & 3989 & 173 & 1359518 \\ 
24.75 & 299797 & 764319 & 549267 & 89454 & 5393 & 291 & 1708521 \\ 
\hline\\[-8pt]
Total & 1416338 & 3476223 & 2389189 & 359629 & 16003 & 640 & 7658021 \\ 
\hline
 \end{tabular}
  \end{center}
\caption{Predicted differential  quasar counts over $15.5<g<25$ and $0<z<6$ for a survey covering 10,000 deg$^2$, based on our best-fit PLE or PLE+LEDE luminosity function model. Bins are centered on the indicated magnitude and redshift values. The ranges in each bin are $\Delta g=0.5$ and $\Delta z = 1$.}
\label{tab:nb_g}
\end{table*}

\begin{table*}[htbp]
\begin{center}
\begin{tabular}{crrrrrrr}
\hline\hline\\[-8pt]
\multicolumn{8}{c}{PLE model}\\
$r\setminus z$ & 0.5 & 1.5 & 2.5 & 3.5 & 4.5 & 5.5 & Total \\  \hline\\[-8pt]
15.75 & 54 & 5 & 14 & 0 & 0 & 0 & 73 \\ 
16.25 & 127 & 23 & 41 & 3 & 0 & 0 & 194 \\ 
16.75 & 323 & 104 & 108 & 9 & 0 & 0 & 544 \\ 
17.25 & 860 & 469 & 290 & 25 & 1 & 0 & 1644 \\ 
17.75 & 2316 & 2009 & 793 & 64 & 2 & 0 & 5185 \\ 
18.25 & 5854 & 7439 & 2204 & 167 & 5 & 0 & 15669 \\ 
18.75 & 12619 & 21145 & 5994 & 432 & 12 & 1 & 40203 \\ 
19.25 & 22090 & 44412 & 14675 & 1100 & 31 & 2 & 82311 \\ 
19.75 & 32538 & 72781 & 29834 & 2700 & 80 & 5 & 137938 \\ 
20.25 & 43787 & 102680 & 50203 & 6208 & 201 & 12 & 203090 \\ 
20.75 & 57009 & 136093 & 74188 & 12952 & 502 & 30 & 280774 \\ 
21.25 & 73477 & 176857 & 101326 & 24039 & 1217 & 75 & 376990 \\ 
21.75 & 94414 & 228620 & 132051 & 39635 & 2811 & 189 & 497720 \\ 
22.25 & 121186 & 295216 & 167766 & 58919 & 6038 & 470 & 649595 \\ 
22.75 & 155438 & 381240 & 210616 & 80786 & 11825 & 1144 & 841048 \\ 
23.25 & 199196 & 492527 & 263236 & 104700 & 20944 & 2660 & 1083262 \\ 
23.75 & 254945 & 636610 & 328743 & 131042 & 33694 & 5769 & 1390804 \\ 
\hline\\[-8pt]
Total & 1076235 & 2598227 & 1382081 & 462783 & 77360 & 10356 & 5607043 \\ 

\hline\hline\\[-8pt]
\multicolumn{8}{c}{PLE+LEDE model}\\
$r\setminus z$ & 0.5 & 1.5 & 2.5 & 3.5 & 4.5 & 5.5 & Total \\  \hline\\[-8pt]

\hline\\[-8pt]
15.75 & 74 & 12 & 2 & 0 & 0 & 0 & 88 \\ 
16.25 & 173 & 46 & 8 & 1 & 0 & 0 & 227 \\ 
16.75 & 425 & 173 & 29 & 3 & 0 & 0 & 630 \\ 
17.25 & 1063 & 645 & 109 & 11 & 1 & 0 & 1830 \\ 
17.75 & 2637 & 2318 & 407 & 37 & 3 & 1 & 5404 \\ 
18.25 & 6171 & 7534 & 1476 & 119 & 10 & 3 & 15314 \\ 
18.75 & 12760 & 20293 & 4934 & 380 & 27 & 7 & 38401 \\ 
19.25 & 22256 & 42853 & 14012 & 1146 & 71 & 15 & 80354 \\ 
19.75 & 33079 & 72019 & 31432 & 3068 & 182 & 33 & 139813 \\ 
20.25 & 44377 & 102993 & 55189 & 6846 & 432 & 69 & 209907 \\ 
20.75 & 56787 & 135480 & 81153 & 12487 & 915 & 133 & 286955 \\ 
21.25 & 71385 & 172381 & 108661 & 19287 & 1689 & 232 & 373634 \\ 
21.75 & 89178 & 216865 & 139912 & 26781 & 2731 & 367 & 475834 \\ 
22.25 & 111154 & 271898 & 177682 & 35182 & 3976 & 534 & 600426 \\ 
22.75 & 138390 & 340627 & 224720 & 45094 & 5393 & 732 & 754956 \\ 
23.25 & 172133 & 426753 & 283998 & 57240 & 7015 & 962 & 948101 \\ 
23.75 & 213829 & 534834 & 359048 & 72418 & 8921 & 1234 & 1190284 \\ 
\hline\\[-8pt]
Total & 975871 & 2347722 & 1482773 & 280101 & 31368 & 4322 & 5122156 \\ 
\hline
 \end{tabular}
  \end{center}
\caption{Predicted differential  quasar counts over $15.5<r<24$ and $0<z<6$ for a survey covering 10,000 deg$^2$, based on our best-fit PLE or PLE+LEDE luminosity function model. Bins are centered on the indicated magnitude and redshift values. The ranges in each bin are $\Delta g=0.5$ and $\Delta z = 1$.}
\label{tab:nb_r}
\end{table*}

We can apply our best-fit QLF to  future surveys, such as the third-generation large-scale spectroscopic survey DESI that aims to observe quasars to a limiting magnitude $g=23$,  or possibly $r=23$ (in order to recover quasars at $z>3.6$ that are $g$-band dropouts due to the absorption of their flux by  Lyman-$\alpha$ absorbers along the line of sight). 
Expected quasar densities at DESI depth for our two models  are given in the last two sections of Tab.~\ref{tab:countsDESI}. The $0.9<z<2.1$ redshift range is where quasars are currently used as direct tracers of dark matter, and the $z>2.1$ regime is where quasars are used to probe dark matter though the Lyman-$\alpha$ forest. The $r<23$ limit produces slightly higher estimates than at $g<23$, because quasars have  positive $g-r$ values  at all redshift ($g-r$ in 0.1 -- 0.5 for $z<3.5$, and increasing at higher redshift).
Despite the differences between the PLE and the PLE+LEDE models that are visible in Fig.~\ref{fig:clum} or in the predicted counts of Tabs.~\ref{tab:nb_g} and \ref{tab:nb_r} for faint magnitudes in particular, the density of quasars predicted by either model over the redshift and magnitude ranges of next-generation surveys are in excellent agreement, at the 1.1\% level for $g<23$ and the 2.5\% level for $r<23$ in total quasar counts.


\section{Conclusions}

The extended Baryon Oscillation Spectroscopic Survey (eBOSS), part of the fourth iteration of the Sloan Digital Sky Survey, has an extensive spectroscopic quasar program that combines several  selection techniques. Algorithms using the  variability  of quasar luminosity with time have been shown to be highly  efficient to obtain large and complete samples of quasars. 

Here we  present the use in eBOSS of time-domain variability, and  focus on a specific program in Stripe 82 that led to a sample of 13,876 quasars to $g_{\rm dered}=22.5$ over a 94.5~deg$^2$ region, 1.5 times denser than expected to be obtained over the rest of the eBOSS footprint. 
This variability program  provides a homogeneous and highly complete  sample of quasars that is denser  and of greater depth than  samples that were used in  previous  studies dedicated to QLF measurements such as \citet{Croom2009,Palanque2013a,Ross2013}. Using the data themselves, plus an external control sample of quasars selected with an independent (color-based) technique, we compute  completeness corrections to account for quasar losses in the selection procedure, in the tiling, or in the spectroscopic identification. 
 
We use this sample to measure the QLF in eight redshift bins from 0.68 to 4.0, and over magnitudes ranging from $M_g(z\!=\!2)=-26.60$ to $-21.80$  at low redshift, and from $M_g(z\!=\!2)=-29.00$ to $-26.20$ at high redshift. The data indicate a break at pivot redshift $z_p=2.2$, with a rise in luminosity followed by a steep decline as the redshift increases on either side of  $z_p$. The data are well fit by a double power-law model. We compare two models that we  constrain to be continuous at $z_p$: a quadratic PLE model as in \citet{Palanque2013a}, with bright-end and faint-end slopes allowed to be different on either side of $z_p$, and a PLE + LEDE model as in \citet{Ross2013}, where a simple linear PLE is used for $z<2.2$ and a LEDE with quadratic magnitude evolution is used at $z>2.2$. These models both have ten parameters free to vary, and they fit the measured binned QLF equally well. Our models are in excellent agreement with \citet{Croom2009} at $z<2.2$, and with  \citet{Masters2012} at $z>2.2$. Our two models start to deviate from one another in the highest two redshift bins ($z>3.0$), although they both are in reasonable agreement with other measurements \citep{Siana2008, Masters2012,Ross2013}.

We use our models to predict densities of quasars for future quasar spectroscopic surveys. We predict $266\pm 3~{\rm deg}^{-2}$ quasars to $g<23$, and $304\pm 7~{\rm deg}^{-2}$ quasars to $r<23$, where the estimate is the mean of the two model estimates and the uncertainty is the difference between the estimate from either model and the mean.

\section*{Appendix}
Tab. \ref{tab:LF} provides the binned luminosity function measured in this work  using the model-weighted estimator described in Sec.~\ref{sec:QLF}, and plotted in Fig.~\ref{fig:clum}. The table lists the value of $\log \Phi$ in eight  intervals spanning redshifts from $z=0.68$ to $z=4.00$, and for $\Delta M_g = 0.40$ magnitude bins from $M_g=-29.00$ to $M_g=-20.60$. We also give the number of quasars ($N_Q$) contributing to the LF in each bin, and the uncertainty ($\Delta \log\Phi$). Bins with quasars but no corresponding value of the binned QLF are data points that were removed in the iterative fitting procedure due to large correction factors.
 
 \begin{table*}[h]
  \begin{center}
  \begin{tabular}{c|rcc|rcc|rcc|rcc}
  \hline
    \hline
  $M_g$ & \multicolumn{3}{c|}{$0.68<z<1.06$} &  \multicolumn{3}{c|}{$1.06<z<1.44$}&  \multicolumn{3}{c|}{$1.44<z<1.82$}&  \multicolumn{3}{c}{$1.82<z<2.20$}\\
(bin center)  & $N_Q$& $\log\Phi$&$\Delta\log\Phi$& $N_Q$& $\log\Phi$&$\Delta\log\Phi$& $N_Q$& $\log\Phi$&$\Delta\log\Phi$& $N_Q$& $\log\Phi$&$\Delta\log\Phi$\\
\hline
-28.80 & 0 & - & - & 0 & - & - & 0 & - & - & 0 & - & - \\
-28.40 & 0 & - & - & 0 & - & - & 0 & - & - & 0 & - & - \\
-28.00 & 0 & - & - & 0 & - & - & 0 & - & - & 4 & -7.63 & 0.22 \\
-27.60 & 0 & - & - & 0 & - & - & 8 & -7.32 & 0.16 & 21 & -6.91 & 0.11 \\
-27.20 & 0 & - & - & 1 & -8.20 & 0.44 & 29 & -6.76 & 0.09 & 41 & -6.63 & 0.08 \\
-26.80 & 0 & - & - & 11 & -7.15 & 0.14 & 51 & -6.51 & 0.08 & 83 & -6.33 & 0.07 \\
-26.40 & 1 & -8.13 & 0.44 & 39 & -6.59 & 0.08 & 94 & -6.25 & 0.07 & 119 & -6.17 & 0.06 \\
-26.00 & 15 & -6.94 & 0.12 & 61 & -6.39 & 0.07 & 131 & -6.10 & 0.06 & 178 & -5.99 & 0.06 \\
-25.60 & 36 & -6.53 & 0.09 & 103 & -6.15 & 0.06 & 197 & -5.92 & 0.06 & 287 & -5.77 & 0.05 \\
-25.20 & 55 & -6.32 & 0.08 & 156 & -5.97 & 0.06 & 337 & -5.68 & 0.05 & 300 & -5.73 & 0.05 \\
-24.80 & 90 & -6.09 & 0.07 & 236 & -5.78 & 0.06 & 388 & -5.59 & 0.05 & 385 & -5.60 & 0.05 \\
-24.40 & 165 & -5.81 & 0.06 & 346 & -5.59 & 0.05 & 427 & -5.52 & 0.05 & 439 & -5.50 & 0.05 \\
-24.00 & 182 & -5.76 & 0.06 & 391 & -5.50 & 0.05 & 487 & -5.42 & 0.05 & 438 & -5.41 & 0.05 \\
-23.60 & 304 & -5.51 & 0.05 & 425 & -5.42 & 0.05 & 468 & -5.36 & 0.05 & 246 & -5.49 & 0.06 \\
-23.20 & 314 & -5.45 & 0.05 & 428 & -5.34 & 0.05 & 370 & -5.30 & 0.05 & 15 & - & - \\
-22.80 & 319 & -5.39 & 0.05 & 288 & -5.35 & 0.05 & 34 & - & - & 0 & - & - \\
-22.40 & 243 & -5.45 & 0.06 & 91 & - & - & 0 & - & - & 0 & - & - \\
-22.00 & 161 & -5.46 & 0.06 & 1 & - & - & 0 & - & - & 0 & - & - \\
-21.60 & 58 & - & - & 0 & - & - & 0 & - & - & 0 & - & - \\
-21.20 & 12 & - & - & 0 & - & - & 0 & - & - & 0 & - & - \\
-20.80 & 0 & - & - & 0 & - & - & 0 & - & - & 0 & - & - \\
  \hline
  \hline
$M_g$ & \multicolumn{3}{c|}{$2.20<z<2.60$} &  \multicolumn{3}{c|}{$2.60 <z<3.00$}&  \multicolumn{3}{c|}{$3.00 <z<3.50$}&  \multicolumn{3}{c}{$3.50 <z<4.00$}\\
  (bin center)   & $N_Q$& $\log\Phi$&$\Delta\log\Phi$& $N_Q$& $\log\Phi$&$\Delta\log\Phi$& $N_Q$& $\log\Phi$&$\Delta\log\Phi$& $N_Q$& $\log\Phi$&$\Delta\log\Phi$\\
  \hline
  -28.80 & 0 & - & - & 0 & - & - & 0 & - & - & 1 & -8.34 & 0.44 \\
-28.40 & 2 & -7.98 & 0.31 & 7 & -7.42 & 0.17 & 3 & -7.89 & 0.26 & 1 & -8.35 & 0.44 \\
-28.00 & 6 & -7.49 & 0.18 & 15 & -7.10 & 0.12 & 11 & -7.33 & 0.14 & 1 & -8.36 & 0.44 \\
-27.60 & 25 & -6.88 & 0.10 & 20 & -6.99 & 0.11 & 11 & -7.33 & 0.14 & 0 & - & - \\
-27.20 & 41 & -6.67 & 0.08 & 31 & -6.80 & 0.09 & 13 & -7.26 & 0.13 & 7 & -7.49 & 0.17 \\
-26.80 & 79 & -6.39 & 0.07 & 52 & -6.57 & 0.08 & 41 & -6.75 & 0.08 & 13 & -7.17 & 0.13 \\
-26.40 & 125 & -6.18 & 0.06 & 91 & -6.32 & 0.07 & 61 & -6.55 & 0.07 & 25 & -6.71 & 0.10 \\
-26.00 & 132 & -6.15 & 0.06 & 124 & -6.16 & 0.06 & 85 & -6.37 & 0.07 & 17 & - & - \\
-25.60 & 192 & -5.96 & 0.06 & 151 & -6.05 & 0.06 & 105 & -6.19 & 0.06 & 6 & - & - \\
-25.20 & 267 & -5.79 & 0.06 & 177 & -5.93 & 0.06 & 69 & -6.23 & 0.07 & 2 & - & - \\
-24.80 & 337 & -5.66 & 0.05 & 169 & -5.88 & 0.06 & 53 & -6.10 & 0.08 & 1 & - & - \\
-24.40 & 282 & -5.67 & 0.05 & 140 & -5.81 & 0.06 & 6 & - & - & 0 & - & - \\
-24.00 & 230 & -5.62 & 0.06 & 20 & - & - & 0 & - & - & 0 & - & - \\
-23.60 & 35 & - & - & 0 & - & - & 0 & - & - & 0 & - & - \\
-23.20 & 0 & - & - & 0 & - & - & 0 & - & - & 0 & - & - \\
-22.80 & 0 & - & - & 0 & - & - & 0 & - & - & 0 & - & - \\
-22.40 & 0 & - & - & 0 & - & - & 0 & - & - & 0 & - & - \\
-22.00 & 0 & - & - & 0 & - & - & 0 & - & - & 0 & - & - \\
-21.60 & 0 & - & - & 0 & - & - & 0 & - & - & 0 & - & - \\
-21.20 & 0 & - & - & 0 & - & - & 0 & - & - & 0 & - & - \\
-20.80 & 0 & - & - & 0 & - & - & 0 & - & - & 0 & - & - \\

  \hline
\end{tabular}
  \end{center}
\caption{Binned quasar luminosity function. The 13,876 quasars  were selected with the variability algorithm presented in Sec.~\ref{subsec:ts}, at $g_{\rm dered}<22.5$.  All quasars lie in a   94.5 $\rm deg^2$ region of Stripe 82 fully observed by eBOSS during the first year of the survey.  The corresponding QLF points  are shown in Fig.~\ref{fig:clum}.}
\label{tab:LF}
\end{table*}

\begin{acknowledgements}

Funding for the Sloan Digital Sky Survey IV has been provided by
the Alfred P. Sloan Foundation, the U.S. Department of Energy Office of
Science, and the Participating Institutions. SDSS-IV acknowledges
support and resources from the Center for High-Performance Computing at
the University of Utah. The SDSS web site is www.sdss.org.

SDSS-IV is managed by the Astrophysical Research Consortium for the 
Participating Institutions of the SDSS Collaboration including the 
Brazilian Participation Group, the Carnegie Institution for Science, 
Carnegie Mellon University, the Chilean Participation Group, the French Participation Group, Harvard-Smithsonian Center for Astrophysics, 
Instituto de Astrof\'isica de Canarias, The Johns Hopkins University, 
Kavli Institute for the Physics and Mathematics of the Universe (IPMU) / 
University of Tokyo, Lawrence Berkeley National Laboratory, 
Leibniz Institut f\"ur Astrophysik Potsdam (AIP),  
Max-Planck-Institut f\"ur Astronomie (MPIA Heidelberg), 
Max-Planck-Institut f\"ur Astrophysik (MPA Garching), 
Max-Planck-Institut f\"ur Extraterrestrische Physik (MPE), 
National Astronomical Observatory of China, New Mexico State University, 
New York University, University of Notre Dame, 
Observat\'ario Nacional / MCTI, The Ohio State University, 
Pennsylvania State University, Shanghai Astronomical Observatory, 
United Kingdom Participation Group,
Universidad Nacional Aut\'onoma de M\'exico, University of Arizona, 
University of Colorado Boulder, University of Portsmouth, 
University of Utah, University of Virginia, University of Washington, University of Wisconsin, 
Vanderbilt University, and Yale University.
\\
We acknowledge D. Vilanova for his help in the selection of the quasar targets with the variability algorithm.

\end{acknowledgements}

\bibliographystyle{aa}
\bibliography{biblio}




\end{document}